\documentclass[epj]{svjour}
\usepackage{graphicx}
\usepackage{amsmath}
\usepackage{amssymb}
\def\slaninafigdir{.}
\begin{document}
\title{%
Critical comparison of several order-book models for stock-market fluctuations
}
\titlerunning{%
Order-book models
}%
\author{%
Franti\v{s}ek Slanina
\inst{1}%
\thanks{e-mail: {\tt slanina@fzu.cz
}
}}
\institute{
Institute of Physics,
 Academy of Sciences of the Czech Republic,
 Na~Slovance~2, CZ-18221~Praha,
Czech Republic
and Center for Theoretical Study, Jilsk\'a 1, Prague, Czech Republic
}
%
%
%
%
%
%
%
\abstract{
Far-from-equilibrium models of interacting particles in one dimension
are used as a basis for modelling the stock-market
fluctuations. Particle types and their positions are interpreted as
buy and sel orders placed on a price axis in the order book. We
revisit some modifications of well-known models, starting with the
Bak-Paczuski-Shubik model. We look at the four decades old Stigler 
model and investigate its variants. One of them is the simplified
version of the Genoa artificial market. The list of studied models is
completed by the models of Maslov and Daniels et al. Generically, in
all cases we
compare the return distribution, absolute return autocorrelation and
the value of the Hurst exponent. It turns out that none of the models
reproduces satisfactorily all the empirical data, but the most promising
candidates for further development are the Genoa artificial market and
the Maslov model with moderate order evaporation.
}
\PACS{%
{89.65.-s
}{Social and economic systems
}%
\and
{05.40.-a
}{ Fluctuation phenomena, random processes, noise, and Brownian motion
}%
\and
{02.50.-r
}{ Probability theory, stochastic processes, and statistics
}%
}
\maketitle%
\section{Introduction}
The order book is the central notion in the stock market. People willing
to buy or sell express their desire in well-specified orders and the
authority of the stock exchange logs all the orders in a list, where
they wait until they are either satisfied (executed) or cancelled. The
visible part of the stock market dynamics, i. e. the complex movement
of the price, is rooted in the detailed and mostly invisible processes
happening within the order book. Anyone who wants to study seriously
the stock market fluctuations, must pay attention to the dynamics of
the order book.

\begin{figure*}[t]
\includegraphics[scale=0.95]{%
\slaninafigdir/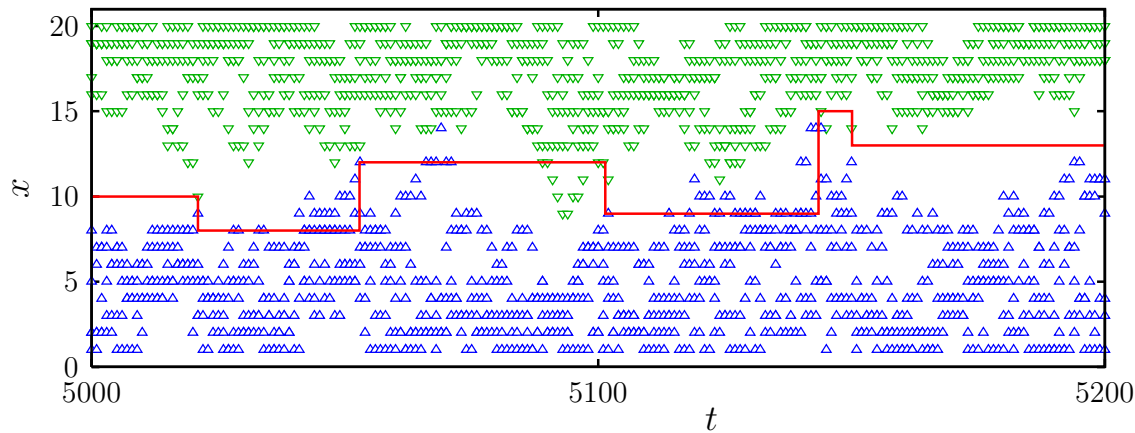}
\caption{
Example of the evolution of the Bak-Paczuski-Shubik model. Triangles
up ($\triangle$) denote positions of bids, triangles down ($\bigtriangledown$)
mark the asks. The full line traces the evolution of the price,
showing
 jumps where transactions occurred. There are $N=5$
particles of each type on the segment of length $L=20$.
}
\label{fig:bps-evol-1-40000-20-5-0}
\end{figure*}

There are several reasons why physicists may and should embark on such
study. First, the discipline of Econophysics is now established and
accepted with decent respect within the Physics community
\cite{an_ar_pi_88,ma_sta_99,bou_pot_00,schweitzer_02}. But even if
the study of economic phenomena by the tools of physics were a bare
empty bubble (which is {\it not!}, the author believes) to be broken
into pieces, the study of the order
book itself may remain one of the shards of value.  (Another one may
be the Minority Game \cite{cha_zha_97}.) Indeed, the second
motivation to spend some effort here is that the order book is a
genuinely one-dimensional non-equilibrium system with complex
dynamics. It abounds with rich phenomena and poses a serious
intellectual challenge, which may provoke development of new tools in
one-dimensional non-equilibrium physics.

The most simplified view of an order book may be the following. The
orders are  
immobile particles of two kinds, $A$ (for asks, i. e. orders to
sell), and $B$ (for bids, i. e. orders to buy), residing on a line of
price (or logarithm of price, if more convenient). All bids are always
on the left of all asks. The actual price lies somewhere between (and
included) the highest bid and the lowest ask. The interval between the
two is the spread and it is one of the key quantities observed in the
order book. Besides these limit orders, waiting for the future in the
order book, also market orders arrive, which buy or sell immediately
at any price available in the market. Thus, the market orders provide
liquidity. 

As we already said, the tip of the order-book iceberg is the
price. All order-book models must be confronted with what is known
about the price fluctuations. These {\it stylised facts} are now very
well established
\cite{ma_sta_95,li_ci_me_pe_sta_97,go_me_am_sta_98,ple_gop_am_mey_sta_99}.  
To quote here only those which we shall be faced
later,  the price movements are generically
characterised by a power-law tail in return distribution, with exponent
$1+\alpha\simeq 4$,
power-law autocorrelation of volatility, with exponent ranging between
$0.3$ to $0.5$, anomalous Hurst
exponent $H\simeq 2/3$,
measured either directly in the so-called Hurst plot, or as a
by-product of another essential feature of the price fluctuations,
which is the scaling. It must be noted, though, that the scaling holds
satisfactorily only for not too long time separations. At larger
times, the gradual crossover to Gaussian shape of return distribution
is observed. This feature is well reproduced in multifractal
stochastic models (from many works in this direction see
e. g. \cite{va_au_98,bac_del_muz_01,eis_ker_04,liu_dim_lux_07}). 
However, we must state from the beginning, that explanation
of multifractality and other subtle features of the stock-market
fluctuations \cite{bou_mat_pot_01,lil_mik_far_05}, goes beyond the
scope of this paper.  

Let us mention at least some of the special
features found empirically in order books. The literature is indeed
very ample
\cite{bia_hil_spa_95,mas_mil_01,cha_sti_01,cha_sti_03,gab_gop_ple_sta_03,ple_gop_gab_sta_04,ple_gop_sta_05,rosenow_02,web_ros_03,web_ros_04,bou_mez_pot_02,pot_bou_02,bou_gef_pot_wya_03,bou_koc_pot_04,wya_bou_koc_pot_vet_06,zov_far_02,lil_far_man_02,lil_far_man_03,far_gil_lil_mik_sen_03,far_pat_zov_03,pon_lil_man_06,far_zam_06,far_ger_lil_mik_06,lillo_06,sca_kai_kir_hub_ted_06,far_lil_04}.
The first thing we may ask is the average order book profile,
i. e. the average number of orders existing in given moment at given
distance form the current price. It was found that it has sharp
maximum very close to, but away from, the price
\cite{cha_sti_01,bou_mez_pot_02,pot_bou_02}. The decrease at large
distances seems to be a power law with exponent $\simeq 2$
\cite{bou_mez_pot_02,pot_bou_02}, but the form of the increase between
the price and the peak is not so clear.

Related information is contained in the price impact function, which
says how much the price moves when an order of a specific volume
arrives. In first approximation, we consider the virtual impact
function, obtained by simple integration of the order book profile
from the current price to the new, shifted price. Beyond the maximum,
the profile decreases and therefore the virtual impact is a convex
function \cite{mas_mil_01,cha_sti_01,web_ros_03}. The striking
surprise in the empirical study of order books is, that the actual
price impact is much smaller, and moreover, it is a concave, rather
than convex, function of volume \cite{web_ros_03}. The form of the
price impact was studied intensively 
\cite{gab_gop_ple_sta_03,lil_far_man_02,lil_far_man_03,far_gil_lil_mik_sen_03,far_pat_zov_03,pon_lil_man_06,far_zam_06,far_ger_lil_mik_06}, 
yet a controversy
persist, whether it can be better fitted on a square root (a qualitative
theoretical argument for this fit can be found in \cite{zhang_99}), a power
with exponent $<0.5$ or on a logarithm. 

The incoming orders have various volumes and it turns out that they
 are power-law distributed \cite{mas_mil_01}. For the market orders,
 the exponent is $\simeq 1.4$, while for the limit orders it has
 higher value $\simeq 2$. The limit orders are deposited at
 various distances from the current price and also here the
 distribution follows a power law
 \cite{bou_mez_pot_02,pot_bou_02,zov_far_02,lillo_06}, although the
 value of the exponent reported differs rather widely ($\simeq 1.5$ to
 $\simeq 2.5$) from one study
 to another. The limit orders are eventually either satisfied or
 cancelled. The 
time they spend within the order book is again power-law distributed 
\cite{cha_sti_01,cha_sti_03,lo_mac_zha_02}
with exponent $\simeq 2.1$ for cancellations and $\simeq 1.5$ for
 satisfactions. 

There were attempts to explain some of the properties of price
fluctuations as direct consequences of the empirically found
statistics of order books. In
Refs. \cite{gab_gop_ple_sta_03,gab_gop_ple_sta_03a}  the power-law
tail in return distribution is related to the specific square-root form of 
the impact function combined with power-law distribution of order
volumes. On the other hand, Ref. \cite{far_gil_lil_mik_sen_03} shows
that the distribution of returns copies the distribution of first gap
(the distance between best and second best order - where ``best''
means ``lowest'' for asks and ``highest'' for bids). It was also found
that the width of the spread is distributed as power law, with
exponent $\simeq 4$ \cite{ple_gop_sta_05}, which is essentially the
same value as the 
exponent for the distribution of returns. The discussion remained
somewhat open, \cite{ple_gop_gab_sta_04,far_lil_04}, but we believe
that the properties of the price fluctuations cannot be deduced entirely
from the statistics of the order book. For example the difference
between the virtual and actual price impact suggests that the order
book reacts quickly to incoming orders and reorganises itself
accordingly. Therefore, without detailed {\it dynamical} information
on the movements deep inside the book we cannot hope for explanation
of the {\it dynamics} of the price.

\section{Existing models}

There is no space here for an exhaustive review of the order-book
modelling, not to speak of other types of stock-market models. 
We select here only a few models we shall build upon in the
later sections and quote only a part of the literature.  We apologise
for unavoidable omissions, not due to underestimation of the work of
others, but dictated by reasonable brevity of this study.

\subsection{Stigler}

To our best knowledge, the first numerical model of the order book and
the first computer simulation ever in economics was the work of
Stigler \cite{stigler_64}. The model is strikingly simple. There are
only limit orders of unit volume and they are supplied randomly into the book 
within a fixed allowed
interval of price. If the new order is e. g. a bid and there is an ask
at lower price, then the bid is matched with the lowest ask and both
of them are removed. If the bid falls lower than the lowest ask, it 
is stored in the book and waits there.

From this example we understand, why the order-book models are often called
 ``zero-intelligence''
 models. Indeed, there is no space for strategic choice of the agents
 and the people may be very well replaced by random number generators.
It is interesting to note that experiments with human versus machine
 trading were performed \cite{god_sun_93}, which found as much
 efficiency in ``zero-intelligence'' machines as in ``rational''
 people (graduate students of business).

\begin{figure}[t]
\includegraphics[scale=0.95]{%
\slaninafigdir/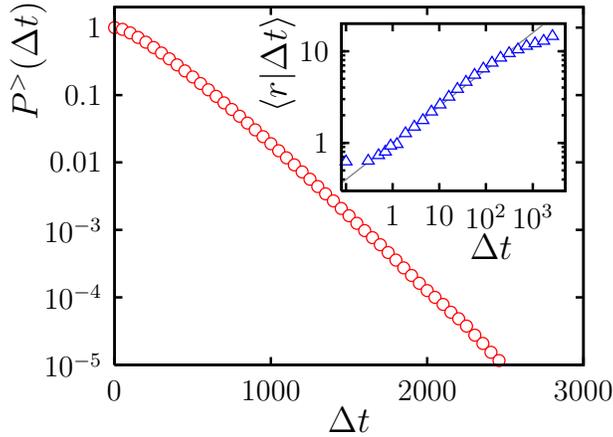}
\caption{
Distribution of inter-event times in BPS model. On the 
segment of length $L=500$, there are  $N=200$  particles of each
kind. In the inset, average return occurring after waiting time
$\Delta t$, for the same values of $L$ and $N$. The line is the power
$\propto (\Delta t)^{0.4}$.}
\label{fig:bps-interevent-time-distr}
\end{figure}

\subsection{Bak, Paczuski, and Shubik}

Another model, very simple to formulate but difficult to solve, was
introduced by Bak, Paczuski, and Shubik (BPS) \cite{ba_pa_shu_97}.
On a line
representing the price axis, two kinds of particles are placed. The
first kind, denoted $A$ (ask),  corresponds to sell orders, while the
second, $B$ (bid), corresponds to buy orders. The position of the
particle is the price at which the order is to be satisfied.
A trade can occur only when two particles of opposite type
meet. If that happens, the orders are satisfied and the particles are removed
from the system. This can be described as annihilation reaction
$A+B\to\emptyset$. It is evident that all $B$ particles must lie on
the left with respect to all $A$ particles.
 The particles diffuse freely and in order to keep
their concentration constant on average, new orders are inserted from
the left ($B$ type) and from the right ($A$ type). The whole picture
of this order-book model is therefore identical to the
two-species diffusion-annihilation process. The changes in the price
are mapped on the movement of the reaction front. 

Many analytical results are known for this model. Most importantly,
the Hurst exponent can be calculated exactly
\cite{krapivsky_95,bar_how_car_96,eli_ko_98,ta_tia_99} 
and the result is $H=1/4$. This value is
well below the empirically established value $H\simeq 2/3$. 

Several modifications of the bare reaction-diffusion process were
introduced \cite{ba_pa_shu_97} to remedy some of the shortcomings of
the model. The simplest one is to postulate a drift of articles
towards the current price. This feature mimics the fact that in real
order books the orders are placed close to the current price. It also
suppresses the rather unnatural assumption of free diffusion of
orders. However, the measured Hurst exponent remains to be $H=1/4$ as
before. 

\begin{figure}[t]
\includegraphics[scale=0.95]{%
\slaninafigdir/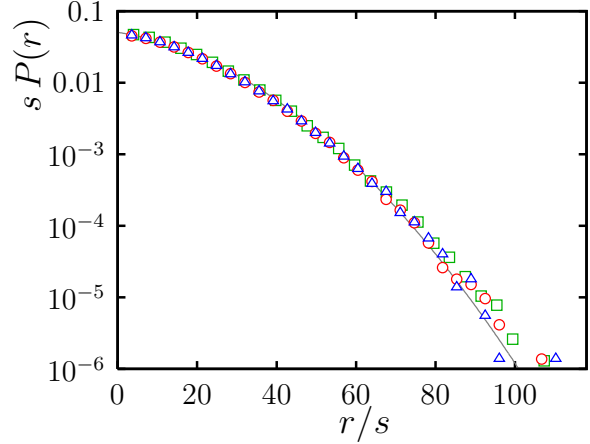}
\caption{
Distribution of one-transaction returns in  BPS model, rescaled
  by the factor $s=N^{1/2}L^{-1/4}$.
The parameters are $L=250$, $N=50$ ($\triangle$); $L=500$, $N=200$
  ({\Large $\circ$}); $L=250$, $N=250$ ($\Box$).
The line is the dependence $\propto
\exp\big(-r/(50s)-(r/(34s))^2\big)$. }
\label{fig:bps-hist}
\end{figure}

\begin{figure}[t]
\includegraphics[scale=0.95]{%
\slaninafigdir/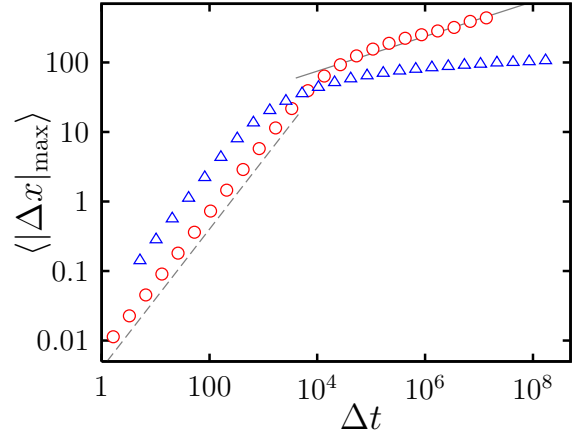}
\caption{
Hurst plot for BPS model. The parameters are 
 $L=2\cdot 10^4$,  $N=2\cdot
10^4$ ({\large $\circ$}), and 
$L=250$, $N=50$
($\triangle$).
The dashed  line is the dependence $\propto \Delta t$, while solid
line is $\propto (\Delta t)^{1/4}$.
}
\label{fig:bps-hurst}
\end{figure}

\begin{figure*}[t]
\includegraphics[scale=0.95]{%
\slaninafigdir/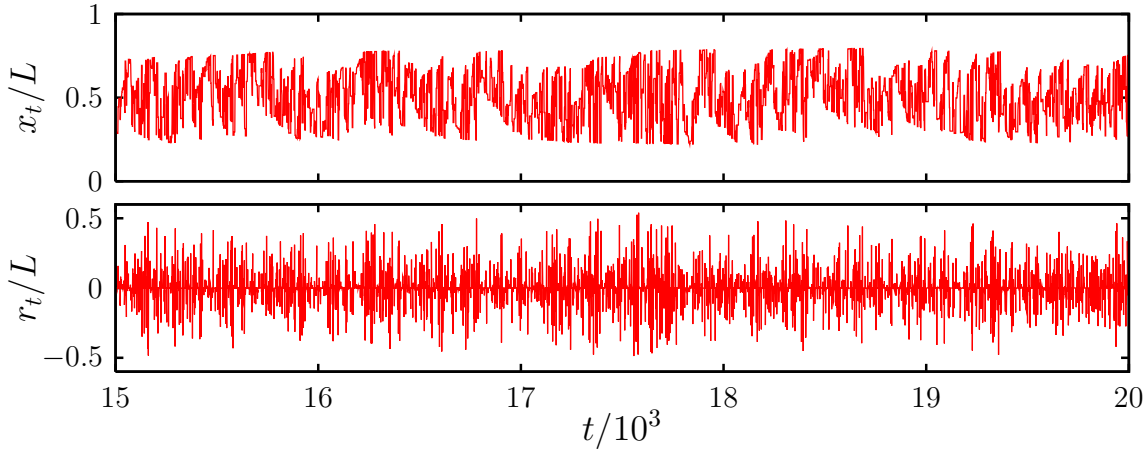}
\caption{
Example of the evolution of the Stigler model. In the upper panel,
time dependence of the actual price; in the lower panel, one-step
returns. On the segment of length $L=5000$ there are at most $N=5000$
orders. 
}
\label{fig:stigler-timeseries}
\end{figure*}

More important modification consists in a kind of ``urn'' process. The 
new orders are placed close to already existing ones, thus
mimicking certain level of ``copying'' or ``herding'' mechanism, which
is surely present in the real-world price dynamics. In this case the
Hurst exponent is higher and in fact very close to the random walk
value, $H\simeq 1/2$. 

The diffusion
constant of the orders can also be coupled to the past volatility,
introducing a positive feedback effect. This way the Hurst
exponent can be enhanced up to the level consistent with the empirical
value. In this case, scaling was observed in the distribution of
returns with Hurst exponent $H\simeq 0.65$.

\subsection{Genoa market model}

The diffusion of orders contradicts reality. Indeed, orders can be
placed into the order book, and later either cancelled or satisfied,
but change in price is very uncommon.  It is therefore wise to return
back to Stigler's immobile orders but to make his model more realistic.

Rather involved modification of the Stigler model appeared much later
under the name of Genoa artificial market 
\cite{rab_cin_foc_mar_01,cin_foc_mar_rab_03,rab_cin_05,rab_cin_foc_mar_05,rab_cin_dos_foc_mar_05,cin_foc_pon_rab_sca_06}.
The model contains many ingredients and is therefore very plastic. 

Again, there are only limit orders and the liquidity is assured by
non-empty intersection of intervals, where the bids and asks,
respectively, are
deposited. In practical implementation, the probability of order
placement was Gaussian, with the centre shifted slightly above the
current price for asks and slightly below for the bids. The width of
the Gaussians was also related to the past volatility, thus
introducing a feedback. Note that essentially the same feedback was
introduced already in the BPS model. 
The price of the contract was calculated
according to demand-offer balance. 
There was also a herding of agents in play, in
the spirit of the Cont-Bouchaud model \cite{co_bou_97}. The main
result to interest us here was the power-law tail of the return
distribution, with very realistic value of the exponent. However, it
was not at all clear which of the many ingredients of the model is
responsible for the appearance of the power-law tail.

\subsection{Maslov model}

To appreciate the crucial role of the market orders, Maslov introduced
a model \cite{maslov_99}, in which the bids are deposited always on
the left and asks on the right from the current price. The limit
orders never meet each other. The execution
of the orders is mediated by the market orders, annihilating the
highest bid or lowest ask, depending on the type of the market order.  

The Maslov model has several appealing features. Especially, the
return distribution characterised by exponent $1+\alpha\simeq 3$ seems 
to be close to the empirically found power
law. The scaling in return distribution is
clearly seen as well as the volatility clustering 
manifested by power-law decay of
the autocorrelation of absolute returns. However, the Hurst exponent
is $1/4$ as in the BPS model, which is  bad news. Maslov model was
treated analytically in a kind of mean-field approximation
\cite{slanina_01}. Unfortunately, the exponent $\alpha=1$ found there 
disagrees with the simulations. Later, the reason for this difference
was identified in the assumption of uniform density of orders on
either of the sides of the price. Taking the density zero at the
current price
and linearly
increasing on both the ask and bid side, the exponent becomes
$\alpha=2$, in agreement with the numerics \cite{maslov_private}.

\subsection{Models with uniform deposition}

The Maslov model is still very idealised. The most important difference from
real situation is the absence of cancellations. In real order books the
orders can be scratched, if their owners think that they waited too
long for their patience. The group of Farmer and others introduced
several variants of models with cancellation (``evaporation'') of
orders
\cite{smi_far_gil_kri_02,dan_far_ior_smi_02,dan_far_gil_ior_smi_03,ior_dan_far_gil_kri_smi_03}. 
Another fundamental feature which makes these models different from
the Maslov model is that the orders are deposited uniformly within
their allowed range, i. e. bids from the current price downwards up
to a prescribed lower bound and equivalently for the asks. 

The order book profile, price impact and many related properties were
studied very thoroughly and their dependence on the rates of thee
processes involved was clarified. An important step forward was the
analytical study performed in \cite{smi_far_gil_kri_02}. Two
complementary ``mean-field'' approaches were applied, achieving quite
good agreement with the simulations. The first approach calculates the
average density of orders as a continuous function, neglecting the
fluctuations. The other approach represents the state of the order
book by intervals between individual orders, assuming that at most one
order can be present on one site (a kind of exclusion
principle). The approximation consists in neglecting the correlations
between the lengths of the intervals. 

\begin{figure}[t]
\includegraphics[scale=0.95]{%
\slaninafigdir/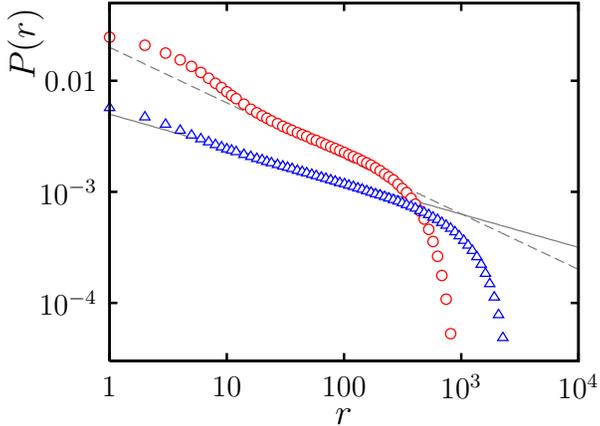}
\caption{
Distribution of one-step returns for Stigler model with $L=5000$ and $N=5000$
($\triangle$) and for the free Stigler model with   $N=5000$,
$s=4000$, and $d=10^4$ 
 ({\Large $\circ$}). The lines are power laws $\propto r^{-0.3}$ (solid)
and $\propto r^{-0.5}$ (dashed).
}
\label{fig:stigler-s2-s3-hist}
\end{figure}

This line of research was recently pushed forward in and important
paper  by Mike and Farmer \cite{mik_far_07}. A scheme, which was given
very fitting name ``empirical model'' was proposed, which incorporates 
several basic empirical facts on the order flow dynamics, namely the
distribution of distances, from the best price, where the orders are placed;
the long memory in the signs of the orders; the cancellation
probability, depending on the position of the order. Including there
empirical ingredients into the Farmer model, an excellent agreement
with other empirical findings was observed, including the return and
spread distributions. The importance of that work, at least from our
point of view, consists in observation that the most tangible feature
of the price fluctuation, the return distribution, is in fact a
secondary manifestation of more basic and yet unexplained
features. These are the features which enter the model of
\cite{mik_far_07} as empirical input.

\begin{figure}[t]
\includegraphics[scale=0.95]{%
\slaninafigdir/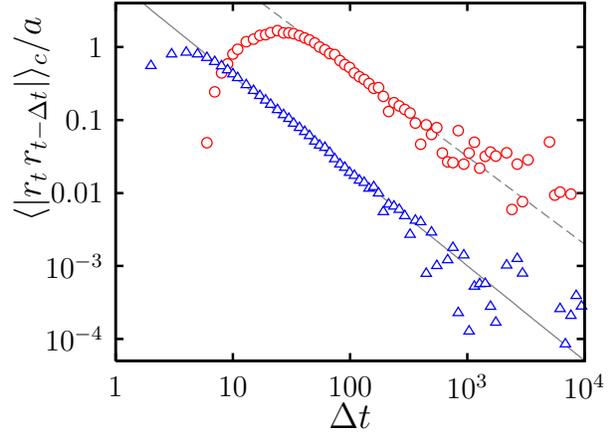}
\caption{
Autocorrelation of absolute returns for the Stigler model with $L=5000$ and $N=5000$
($\triangle$) and for the free Stigler model with  $N=5000$, $s=4000$, and $d=10^4$
 ({\Large $\circ$}). The lines are power laws $\propto (\Delta t)^{-1.3}$ (solid)
and $\propto (\Delta t)^{-1.2}$ (dashed). In order to have all data in the same
frame, we introduced an auxiliary factor $a=10$ ({\Large $\circ$}) and 
$a=10^4$ ($\triangle$).
}
\label{fig:stigler-s2-s3-autocor}
\end{figure}

\begin{figure}[t]
\includegraphics[scale=0.95]{%
\slaninafigdir/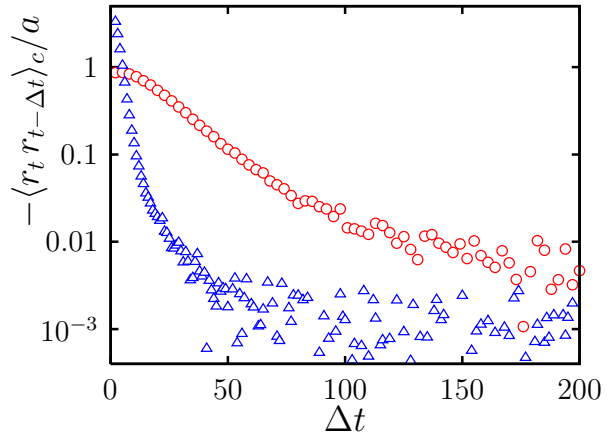}
\caption{
Autocorrelation of  returns for the Stigler model with $L=5000$ and $N=5000$
($\triangle$) and for the free Stigler model with  $N=5000$, $s=4000$, and $d=10^4$
 ({\Large $\circ$}).  In order to have all data in the same
frame, we introduced an auxiliary factor $a=100$ ({\Large $\circ$}) and 
$a=10^4$ ($\triangle$).}
\label{fig:stigler-s2-s3-r-autocor}
\end{figure}

In our work, we address a less ambitious but more fundamental
question. What will be the fluctuation properties of these models
without assuming anything special about order flow? We shall see that
in many aspects the answer is disappointing in the sense that the
results are often far from reality. This means that the inputs of
\cite{mik_far_07} are essential. On the other hand, we can hardly be
satisfied until we detect the causes behind the empirical
ingredients of \cite{mik_far_07}.

\subsection{Other approaches}

A rather phenomenological model was simulated in
\cite{bou_mez_pot_02}. The profile of the order book was successfully 
explained assuming power-law distribution of placement distances from
the current price. 

In fact, the crucial role of the evaporation of orders was first
noticed in the work of Challet and Stinchcombe \cite{cha_sti_01}. The
new limit orders were deposited close to the price, with standard
deviation which was linearly coupled with the width of the spread. The
evaporation caused a clearly visible  crossover from Hurst exponent
$H=1/4$ at short time distances to the random-walk value $H=1/2$
at larger times. This class of models was investigated in depth
subsequently \cite{cha_sti_03,cha_sti_02,stinchcombe_06}. 
In a related development, a version of asymmetric exclusion model
\cite{ha_zha_95} was adapted as an order-book model  
\cite{wil_sch_cha_02}. The two crucial ingredients are the (biased) diffusion
of particles (orders), 
returning somewhat back to the BPS model, and the exclusion
principle, allowing at most one order at one site. It also forbids
``skipping'' of 
particles, so each order represents an obstacle for the diffusion of
others. Price is represented by the particle of a special
type. Mapping to the exactly soluble asymmetric exclusion model gives
the precise value of the Hurst exponent $H=2/3$, nicely coinciding with
reality. One must remember, though, that the price for this result is
the unrealistic assumption of diffusing orders. Moreover, even if we
accepted the view that removal and immediate placement of an order not
far from the original position may be effectively described as
diffusion, why then the particles are not allowed to overtake each
other? We consider that feature very far from reality.

\begin{figure}[t]
\includegraphics[scale=0.95]{%
\slaninafigdir/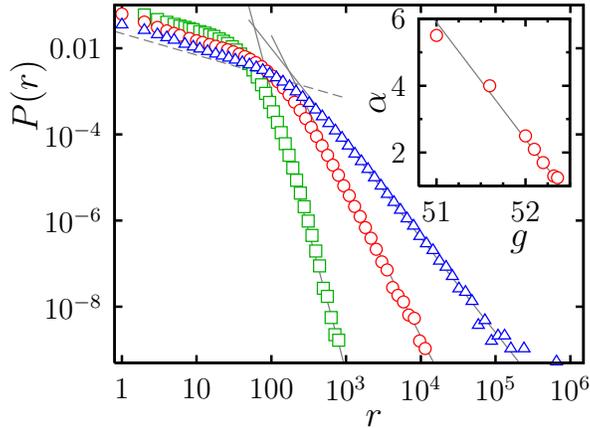}
\caption{
Return distribution in the Genoa market model. Maximum number of
orders is
$N=1000$, width to shift ratio is $b=7$. The feedback factor is $g=51$
($\Box$), $52$ ({\Large $\circ$}), and $52.36$ ($\triangle$). The
three 
solid lines are power laws $\propto r^{-1-\alpha}$ with the exponents (from
left to right) $\alpha=5.5$, $2.5$, and $1.2$.  The dashed line is the power 
$\propto r^{-0.5}$. 
In the inset, the
dependence of the tail exponent $\alpha$ on the feedback factor
$g$. The line is the dependence $(\alpha-1)\propto(52.4-g)$ indicating
that the critical value lies at $g_c\simeq 52.4$.
}
\label{fig:genoa-return}
\end{figure}

Let us only list some other works we consider relevant for order-book
modelling
\cite{osborne_65,kul_ker_01,fra_mar_mat_01,mat_fra_01a,far_josh_00,challet_06}.
Schematic models, like the Interacting Gaps model
\cite{mu_sla_sol_03,svo_sla_07}, 
may also bring some, however limited, insight.
Despite continuing effort of many groups performing empirical analyses
as well as theoretical studies, the true dynamics of the order book is
far from being fully 
understood. On one side, the trading in the stock market 
is much more intricate than mere play of limit and market
orders. There are many more types of them, sometimes rather
complicated. At the same time, it becomes more and more 
evident that assuming ``zero-intelligence'' players misses some
substantial processes under way in the stock market. Strategic
thinking cannot be avoided without essential loss. This brings us close
to our last remark.  All the models mentioned in this section 
are appropriate only to those markets, which operate 
without an official market maker. In presence of a market maker, the
orders do not interact individually, but in smaller or larger
chunks. One is tempted to devise a ``zero-intelligence'' model with a
market maker, but there is perhaps a wiser path to follow. We have in
mind a combination of order-book models with Minority
Game. The latter represents an antipole to ``zero-intelligence''
order-book models and amalgamating the two opposites may prove fruitful.

\begin{figure}[t]
\includegraphics[scale=0.95]{%
\slaninafigdir/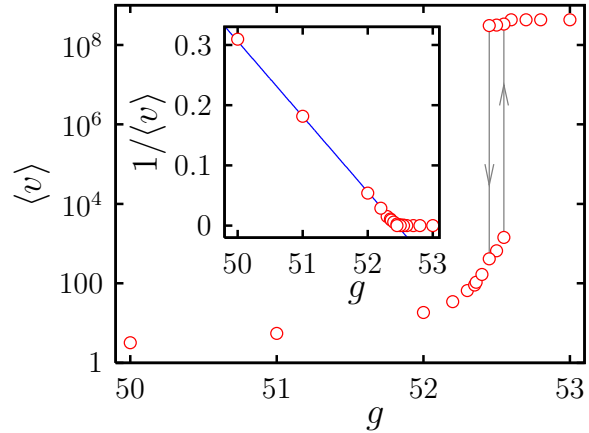}
\caption{
Genoa market model. Dependence of the average volatility on the
feedback factor $g$. The parameters are $N=1000$,  $b=7$. 
The lines with arrows indicate the
hysteresis curve, the false signature of an apparent first-order transition. In the
inset, the same data but plotted differently. The line is the
dependence $\propto(52.4-g)$, suggesting the critical value $g_c\simeq 52.4$.
}
\label{fig:genoa-volatility}
\end{figure}

\begin{figure}[t]
\includegraphics[scale=0.95]{%
\slaninafigdir/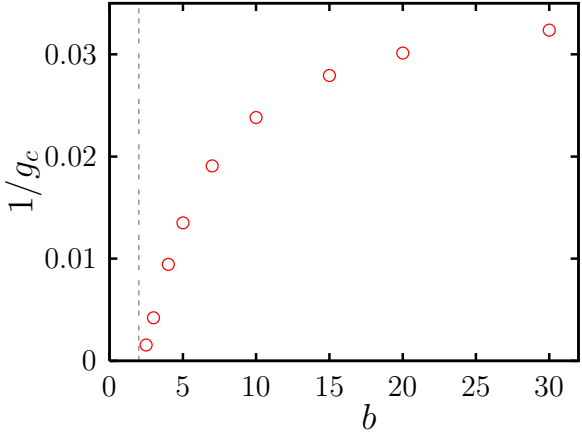}
\caption{
Phase diagram of the Genoa market model for $N=1000$. Inverse of the critical value $g_c$
of the feedback factor, deduced from the simulations, depends on the
width to shift ratio $b$. The phase transition is absent in the
(trivial) region $b<2$, indicated by dashed line.
}
\label{fig:genoa-phasediagram}
\end{figure}

In this work we shall not go thus far. Our aim is rather to clarify
the dark places in the ensemble of existing order-book
models. Performing new simulations for several of these models in
parallel, we hope to shed some light on the the usefulness
 and the limitations of them.

\section{New simulations}

Here we present our new results of numerical simulations of the models
sketched above. Some of the data aim at improving the results already
present in the literature, but mostly we try to clarify aspects not
studied before. We also used the same methodology in analysing the
simulations for all models, in order to make comparable statements 
for each of the models under scrutiny.

\begin{figure}[t]
\includegraphics[scale=0.95]{%
\slaninafigdir/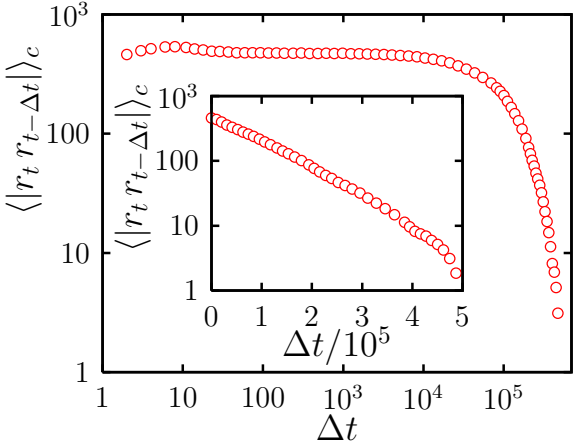}
\caption{
Autocorrelation of absolute returns in the Genoa market model, for parameters
$N=1000$, $b=7$, $g=52$. In the inset, the same data are plotted in linear-logarithmic scale.
}
\label{fig:stigler-s4-autocor}
\end{figure}

\subsection{Bak-Paczuski-Shubik model}

The first model to study is the Bak-Paczuski-Shubik (BPS) model. As we
already explained, we have two types of diffusing particles, called
$A$ and $B$. There are $N$ particles of each type, i. e. total $2N$
particles placed at the segment of length $L$. The particles can
occupy integer positions from the set $\{1,2,\ldots,L\}$. In one
update step we choose one particle and change its position as
$c'_i=c_i\pm 1$ (there is no bias, so both signs of the change have
the same probability), on condition that the new position stays
within the allowed interval, $1\le c'_i\le L$.  
We use the convention that the time advances by $1/(2N)$ in
one step. If the new site was empty or there 
was already another particle of the same type at the
new position, nothing more happens an the update is completed. We set
$c_i(t+1/(2N))=c'_i$ and $c_k(t+1/(2N))=c_k(t)$, $k\ne i$ On the
other hand, if the new site is occupied by a particle of opposite
type, say, particle $j$, so that $c_j(t)=c'_i$, then the two
particles annihilate. To keep the number pf particles constant, we
immediately supply two new particles at opposite edges of the allowed
segment. E. g. if $i$ was type $B$ and $j$ was type $A$, the update is
$c_i(t+1/(2N))=1$, $c_j(t+1/(2N))=L$ and $c_k(t+1/(2N))=c_k(t)$, $k\ne
i,j$. 

The annihilation corresponds to an elementary transaction. The price
set in this deal is just the position where the annihilation took
place, $x(t+1/(2N))=c'_i$. If the transaction did not occur, the price
stays unchanged, $x(t+1/(2N))=x(t)$.  This completes the definition of
the variant of the BPS model simulated here.

In Fig. \ref{fig:bps-evol-1-40000-20-5-0} we can see how the typical
configuration of orders evolves in time. There are rather long periods
where the price does not change, but the positions of orders are mixed
substantially. We shall first look at these waiting times between
consecutive trades. In Fig. \ref{fig:bps-interevent-time-distr} we can
see the (cumulative) probability distribution of them. It is evident
that the distribution is exponential, or very close to it, so we can
consider the sequence of trade times at least approximately as Poisson
point process.

\begin{figure}[t]
\includegraphics[scale=0.95]{%
\slaninafigdir/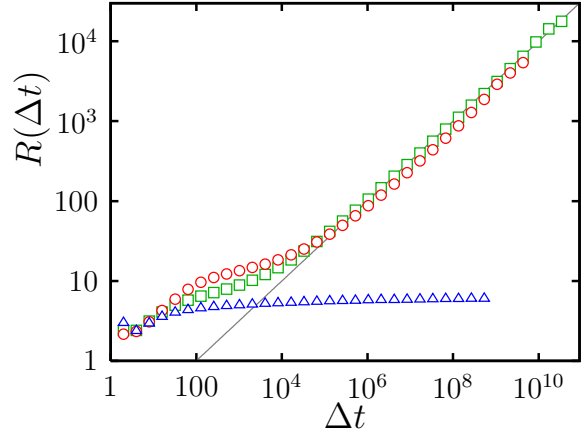}
\caption{
Comparison of Hurst plots for Stigler model with parameters $L=5000$, $N=5000$
($\triangle$), free Stigler model 
with $N=5000$, $s=4000$, $d=10^4$ ({\Large $\circ$}), and Genoa market
model with $N=1000$, $b=7$, $g=51.6$ ($\Box$). The line is the power
$\propto (\Delta t)^{1/2}$.
}
\label{fig:stigler-s5-s6-s7-hurst}
\end{figure}

The most desired quantity is the one-trade return
distribution. If $t_i$ is the time of $i$-th trade, we define 
$r(t_i)=x(t_{i+1})-x(t_i)$ and in Fig. \ref{fig:bps-hist} we plot the
distribution of the absolute returns
$P(r)=\langle\delta(r-|r(t_i)|)\rangle$ 
in stationary state, for
several sizes $L$ and particle numbers $N$. We find that the
distribution collapses onto a single curve when we rescale the data by
the factor 
\begin{equation}
s=N^{1/2}L^{-1/4}\;.
\end{equation}
We then find 
\begin{equation}
P(r)=\frac{1}{s}\,F(\frac{r}{s})
\end{equation}
and the scaling function decays faster than an exponential. The fit
of the type $F(x)\simeq A\exp(-ax-bx^2)$ seems to be fairly satisfactory. 
Evidently, this distribution is very far from the fat tails observed
empirically. It is also interesting to see how the one-trade return
depends on the waiting time before the trade. We measure the
conditional average of the return 
\begin{equation}
\langle r|\Delta t\rangle=
\frac{\sum_i|r(t_i)|\,\delta(t_i-t_{i-1}-\Delta t)}{
\sum_i\delta(t_i-t_{i-1}-\Delta t)}
\end{equation}
and find (see the inset in Fig. \ref{fig:bps-interevent-time-distr})
that it increases slowly as a power law $\langle r|\Delta t\rangle\sim
(\Delta t)^{0.4}$. 

\begin{figure*}[t]
\includegraphics[scale=0.95]{%
\slaninafigdir/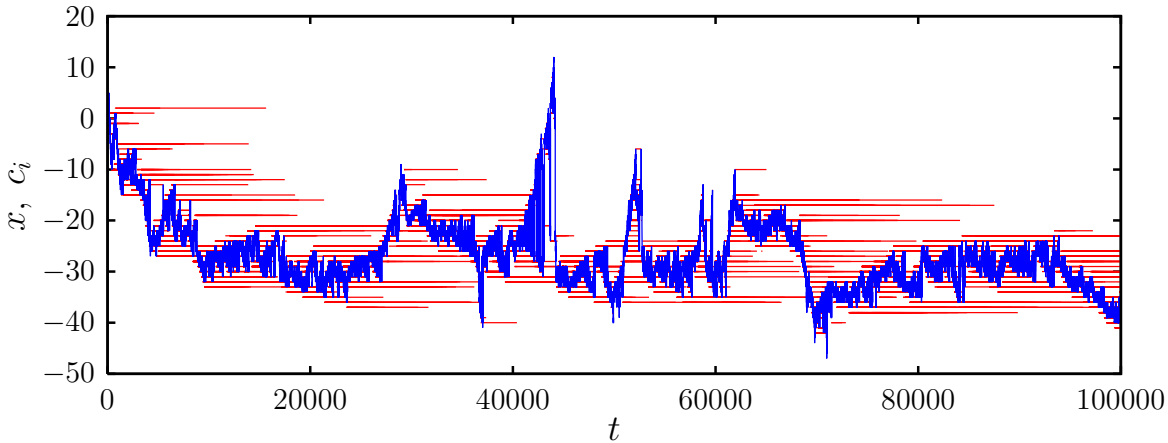}
\caption{
Example of the evolution of the Maslov model with evaporation. Each
segment of a horizontal line corresponds to one order, placed where
the segment starts and executed or evaporated where the segment
ends. The rugged line is the time dependence of the actual price.
Average
number of orders is $\overline{N}=100$ and the probability of  evaporation $q=0.05$.
}
\label{fig:maslovmodel-m5-evol-1-100000-0.05-100}
\end{figure*}

Diffusion of the price is quantified by the Hurst plot. Usually we
calculate the quantity
\begin{equation}
R(\Delta t)=\Bigg\langle\frac{
\max_{t',t''\in(t,t+\Delta t)}\big|x(t')-x(t'')\big|
}{
\sqrt{\big\langle r^2(t')\big\rangle_{t'}
-\big\langle r(t)\big\rangle^2_{t'}}
}\Bigg\rangle_t
\label{eq:hurst-complete}
\end{equation}
where the average $\langle\ldots\rangle_{t'}$ is taken over interval
$t'\in(t,t+\Delta t)$ while the average  $\langle\ldots\rangle_{t}$
extends over
all times. The time-dependent normalisation in the denominator of
(\ref{eq:hurst-complete}) accounts for temporal variations of the
volatility.

However, especially in BPS model the measure (\ref{eq:hurst-complete})
is inconvenient as it does not cover properly the time scales below
the typical waiting time. We use instead a simplified and also
frequently used quantity
\begin{equation}
\big\langle|\Delta x|_\mathrm{max}\big\rangle
=\Big\langle
\max_{t',t''\in(t,t+\Delta t)}\big|x(t')-x(t'')\big|
\Big\rangle_t\;.
\label{eq:hurst-simplified}
\end{equation}
Both (\ref{eq:hurst-complete}) and (\ref{eq:hurst-simplified}) are
expected to share the same asymptotic behaviour for $\Delta
t\to\infty$, i. e. $R(\Delta t)\sim
\big\langle|\Delta x|_\mathrm{max}\big\rangle\sim (\Delta t)^H$ with
Hurst exponent $H$.

The results for BPS model are shown in Fig. \ref{fig:bps-hurst}. We
can appreciate there how difficult it is to actually observe the value
$H=1/4$ predicted by the theory. Relatively long ``short-time'' regime 
seen in Fig. \ref{fig:bps-hurst} is characterised by $H=1$, which
corresponds to ballistic, rather than diffusive, movement of the
price. In this regime, 
the time scale is shorter than the average inter-event time, so
there is typically at most one transaction. The transaction times
follow approximately the Poisson point process, so the probability that one
transaction occur during time $\Delta t$ is, for short times, 
proportional to $\Delta t$. Assuming that the price change, if it
occurs,
 has certain
typical size, the scale of the average price change
should be also proportional to $\Delta t$. Hence the ballistic
behaviour $H=1$ seen in the Hurst plot. Note, however, that this
argument needs some refinement, because, as we have seen in
Fig. \ref{fig:bps-interevent-time-distr}, longer waiting times imply
larger price jumps afterwards. Nevertheless, we believe that the
general line of the argument is true.

The behaviour changes when $\Delta t$ becomes comparable to the average
inter-event time. The most often encountered result is represented by
triangles in Fig. \ref{fig:bps-hurst}. At scales larger than the
average inter-event time the quantity 
$\big\langle|\Delta x|_\mathrm{max}\big\rangle$ 
saturates, yielding $H=0$. It is easy to understand why it must be
so. If the density of particles is large enough, the configuration of
the order book can be described by average concentrations $\rho_A(y)$
and $\rho_B(y)$ of particles $A$ and $B$, respectively. The variable
$y\in(0,L)$ measures the position on the price axis. It is easy to
find that neglecting the fluctuations in the order density 
the solution of the BPS model trivialises into 
$\rho_B(y)=\frac{8N}{L^2}(L/2-y)\theta(L/2-y)$, 
$\rho_A(y)=\frac{8N}{L^2}(y-L/2)\theta(y-L/2)$.
So, in absence of fluctuations the price is pinned in the exact middle
of the allowed interval. This is just the saturation regime $H=0$.

To see the theoretically predicted Hurst exponent $H=1/4$ we must find a
time window between the ballistic and pinned regime. This is often
very narrow, if it exists at all, as testified in
Fig. \ref{fig:bps-hurst}
by the data for $L=250$ and $N=50$. Only for large enough size with
small enough density of orders the fluctuation regime $H=1/4$ is
observable. (Note that in the finite-size analysis the number of
orders must scale as $N\propto L^2$ with the length of the allowed
interval.) 
In Fig. \ref{fig:bps-hurst} we can see an example for
$L=N=2\cdot 10^4$, where such time window is visible. 

The difficulty to observe the desired regime in BPS model contrasts
with the way the exponent $H=1/4$ was derived analytically
\cite{krapivsky_95,bar_how_car_96}. In these works the two reactants
occupy initially the positive and negative half-lines,
respectively. Then, they are let to diffuse and react. Annihilated
particles are not replaced. Therefore, the reaction front spreads out
indefinitely and we can observe  a well defined long-time regime
characterised by the exponent $H=1/4$. (There is also a logarithmic
factor there, but we neglect it in this discussion.) On the contrary,
in BPS the long-time regime has always $H=0$. 

\begin{figure}[t]
\includegraphics[scale=0.95]{%
\slaninafigdir/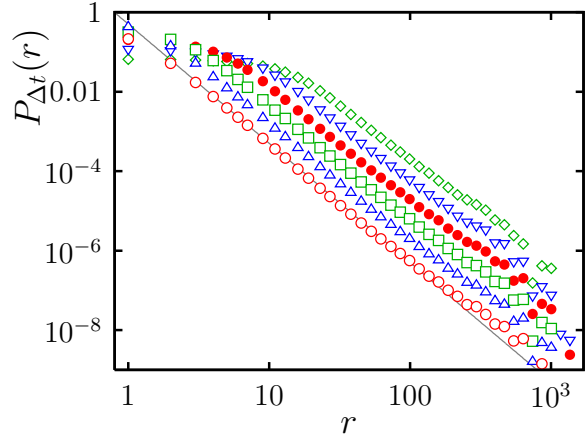}
\caption{
Distribution of returns in the
Maslov model without evaporation, at time lags
 $\Delta t=1$ ({\Large $\circ$}), $10$ ($\triangle$), $100$
($\Box$), $10^3$ ({\Large $\bullet$}), $10^4$
 ($\bigtriangledown$), and $10^5$ ({\Large $\diamond$}).
The line is the power $\propto r^{-3}$. 
}
\label{fig:maslovmodel-m5-hist-0-allags}
\end{figure}

\subsection{Stigler model and its free variant}

In Stigler model, we have again the allowed price range
$\{1,2,\ldots,L\}$, where the orders can be placed. There can be at
most $N$ orders total. If, at time $t$, there is still the order
deposited at time $t-N$, it is removed. Then, we deposit a new
order. We decide whether it will be a bid or an ask (with equal
probability) and choose randomly, with uniform distribution, its
position within the allowed price range. A transaction may follow. If
the new order is e. g. a bid placed at position $c_t$ and the lowest
ask is at position $c_A\le c_t$, then the new price is set to
$x_t=c_A$ and both the new bid at $c_t$ and the old lowest ask at
$c_A$ are removed. 
If  $c_A > c_t$, the price does not change, $x_t=x_{t-1}$ and the new
bid stays in the order book. (Symmetrically it holds for depositing an
ask.) 

In Fig. \ref{fig:stigler-timeseries}
we show an example of the typical time sequence of price $x_t$ and
one-step returns $r_t=x_t-x_{t-1}$. Qualitatively, we can guess that
the fluctuations are far from Gaussian, i. e. returns
will not obey the normal distribution. Indeed, we can see in Fig. 
\ref{fig:stigler-s2-s3-hist} that for several decades 
the distribution falls off slowly as
a power with small exponent, $P(r)\sim r^{-0.3}$ and then it is
sharply cut off. Indeed, the cutoff comes 
from the natural bound $|r_t|<L$.

In the time series in Fig. \ref{fig:stigler-timeseries}
we can also glimpse the volatility clustering. To measure it
quantitatively, we plot in Fig. \ref{fig:stigler-s2-s3-autocor}
the autocorrelation of absolute returns
\begin{equation}
\langle| r_t\,r_{t-\Delta t}|\rangle_c=
\langle| r_t\,r_{t-\Delta t}|\rangle-\langle|
r_t|\rangle\langle|r_{t-\Delta t}|\rangle \;.
\end{equation}
It decays as a power, but with rather large exponent, $\langle|
r_t\,r_{t-\Delta t}|\rangle_c\sim (\Delta t)^{-1.3}$. On the other
hand, the returns themselves are only short-time negatively correlated
with exponential decay, as can be seen in
Fig. \ref{fig:stigler-s2-s3-r-autocor}. 

These findings show that Stigler model is not a very good candidate
model for 
explaining the empirical facts. However, it may well serve as a
starting point 
for successful construction of better models. The first limitation we
must remove is the fixed range of prices from $1$ to $L$. A severe
consequence of this limitation is the saturation seen in the Hurst plot
(Fig. \ref{fig:stigler-s5-s6-s7-hurst}). In long time regime, the
Hurst exponent is obviously $H=0$. To cure this problem we introduce a
``free'' variant of the Stigler model. It may be also considered as a
precursor of the Genoa market model, to be studied in the next
section.

\begin{figure}[t]
\includegraphics[scale=0.95]{%
\slaninafigdir/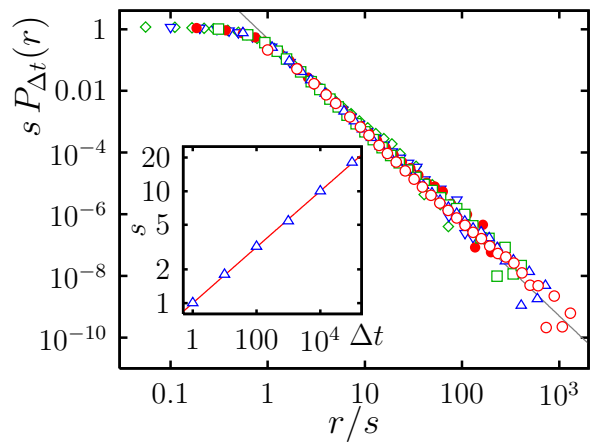}
\caption{
Rescaled distribution of returns in the
Maslov model without evaporation.
The meaning of the symbols is the same as in Fig. \ref{fig:maslovmodel-m5-hist-0-allags}.
The line is the power $\propto r^{-3}$. 
In the inset we plot the dependence of the scaling constant on the
time lag. The line is the power $\propto (\Delta t)^{1/4}$.
}
\label{fig:maslovmodel-m5-hist-0-allags-scaled}
\end{figure}

The price axis is now extended to all integer numbers. Of course, the
position on this axis must be now interpreted as logarithm of price,
rather than price itself. Nonetheless, for brevity we shall speak of
``price'' also in this case. The orders are again deposited randomly
within an allowed range, but now the range depends on the actual
position of the price $x_t$. We introduce two integer 
parameters, the width of
the allowed interval $d$ and the shift $s$ of the interval's centre
with respect to the current price. Denote $c_t$ the order issued at
time $t$. If it is a bid, it is  deposited uniformly
within the range $x_t-s-d/2<c_t\le x-s+d/2$, while for an ask the
range is $x_t+s-d/2\le c_t < x+s+d/2$.  Of course, in order to have
any transactions at all, we must have $d\ge 2s$. As with the Stigler model, the
orders older than $N$ steps are removed.

In spite of the change in the deposition rules, the basic features of
the free Stigler model remain very similar to those of the original
variant. In Fig. \ref{fig:stigler-s2-s3-hist} we can see that the
return distribution exhibits slow power-law decay $P(r)\sim r^{-0.5}$
with a sharp cutoff at large returns. The exponent $\simeq 0.3$ 
is larger than in
the Stigler model, but still remains very much below the empirical
value $\simeq 4$. The autocorrelation of absolute returns (see 
Fig. \ref{fig:stigler-s2-s3-autocor})
decays as a similar power law $\langle|
r_t\,r_{t-\Delta t}|\rangle_c\sim (\Delta t)^{-1.2}$. In addition, a
peak in the autocorrelation function, merely visible in Stigler model,
becomes quite pronounced here and is shifted to larger times, about
$(\Delta t)_\mathrm{peak}\simeq 20$. This indicates some
quasi-periodic pattern in the time series of the volatility, related
probably to a typical waiting time between subsequent trades. Indeed,
we found that the waiting times are exponentially distributed, and for
the parameters of Fig. \ref{fig:stigler-s2-s3-autocor}
the average waiting time is about $\simeq 11$.
As for
the autocorrelation of 
returns, it decays exponentially again, albeit
more slowly, as shown in Fig. \ref{fig:stigler-s2-s3-r-autocor}.

\begin{figure}[t]
\includegraphics[scale=0.95]{%
\slaninafigdir/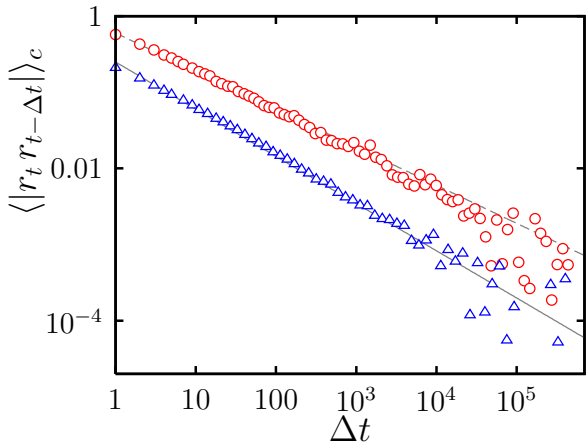}
\caption{
Autocorrelation of absolute returns for the Maslov model without
evaporation ({\Large$\circ$}) and with evaporation probability
$q=0.01$ ($\triangle$).  Average number of orders is $\overline{N}=1000$.
The dashed line is the power $\propto(\Delta t)^{-0.5}$ and the solid
line is $\propto(\Delta t)^{-0.62}$.
}
\label{fig:maslovmodel-m5-autocor-all}
\end{figure}

The main difference observed, compared to the original Stigler model,
is shown in the Hurst plot, Fig. \ref{fig:stigler-s5-s6-s7-hurst}. At
shorter times, there is a tendency to saturation, as in the Stigler
model, but at larger times the purely diffusive regime with $H=1/2$
prevails. We can attribute these results the following
interpretation. The orders present in the order book form a ``bunch''
located somewhere around the current price. Orders too far from the price 
are usually cancelled after their lifetime (equal to $N$)
expires. Hence the localisation around the price. Now, while in the
Stigler model the bunch of orders is imprisoned between $1$ and $L$,
in the free Stigler model the bunch can wander around, following the
price changes. The value $H=1/2$ shows that the movements of the bunch
as a whole can be described as an ordinary random walk.

\subsection{Genoa market model}

Both in original and free Stigler model, the agents behind the scene
have truly zero intelligence. At most, they look at the 
price in this instant and place orders at some distance from it, but
 the distance is not affected neither by the present nor the past
 sequence of prices. However, it is reasonable to expect that the
 agents react to the fluctuations observed in the past. The simplest
 feedback mechanism may be that the distance to place an order is
 proportional to the volatility measured during some time period in
 the past. This idea was already applied in one of the variants of the
 BPS model \cite{ba_pa_shu_97} and lies in the basis of the Genoa 
artificial market
\cite{rab_cin_foc_mar_01}. What we shall call 
``Genoa market model'' from now on, is in fact very reduced version of
the complex simulation scheme of Ref. \cite{rab_cin_foc_mar_01}. We
believe, however, that we retain the most significant ingredients. 

\begin{figure}[t]
\includegraphics[scale=0.95]{%
\slaninafigdir/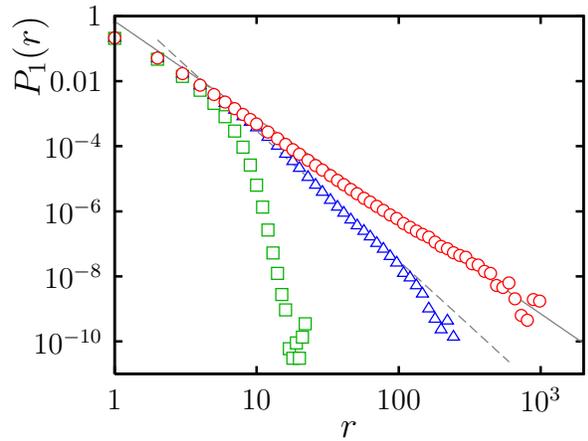}
\caption{
Distribution of one-step returns in the Maslov model with
($\triangle$, $\Box$) and without ({\Large $\circ$}) evaporation. 
The evaporation probability is $q=0.01$ ($\triangle$), $0.05$
($\Box$); the average number of orders is $\overline{N}=1000$. 
The solid line is the power $\propto r^{-3}$, the
dashed line is $\propto r^{-4}$.
}
\label{fig:maslovmodel-m5-hist-all}
\end{figure}

We must first define a convenient measure of instantaneous 
 volatility. Averaging absolute price changes with an exponentially
 decaying kernel 
\begin{equation}
v_t=\lambda\sum_{t'=0}^\infty (1-\lambda)^{t'}|x_{t-t'}-x_{t-t'-1}|\;.
\end{equation}
turns out to be a good choice. We use the value $\lambda=10^{-3}$
throughout the simulations. The orders will be placed on integer
positions within an
interval determined by the width and the shift from actual price, as
in the free Stigler model, but now these two parameters are
time-dependent. Their ratio will be held constant and both will expand
as the volatility $v_t$ will grow. So, the prescription will be
\begin{equation}
\begin{split}
&d_t=\lceil g\, v_t\rceil\\
&s_t=\Big\lfloor\frac{d_t}{b}\Big\rfloor
\label{eq:orderbook-genoa-feedback}
\end{split}
\end{equation}
and the constants $b$  and $g$, besides the maximum number of orders
(i. e. maximum lifetime of an order) $N$ constitute the parameters of
the model. In order that we have any transactions at all, we impose
the bound $b>2$.

The feedback mechanism we apply makes significant difference in all
aspects of the model. Let us look first at the return distribution. In
Fig. \ref{fig:genoa-return} we can see how it changes when we tune the
parameter $g$. Generically, a power-law tail $P(r)\sim r^{-1-\alpha}$ 
develops, with an
exponent strongly depending on $g$. The larger $g$, the smaller the
exponent, until for some critical value $g=g_c$ it approaches the
limit $\alpha=1$. beyond that point, the average returns, i. e. also
the stationary value of the average volatility $v_t$ diverges. This
may be regarded as a kind of phase transition. It is also worth nothing 
that for low returns there is an interval where another power law
holds, with $1+\alpha\simeq 0.5$.  This is the remainder of the
behaviour characteristic for the free Stigler model, the parent of the
Genoa stock market.

\begin{figure}[t]
\includegraphics[scale=0.95]{%
\slaninafigdir/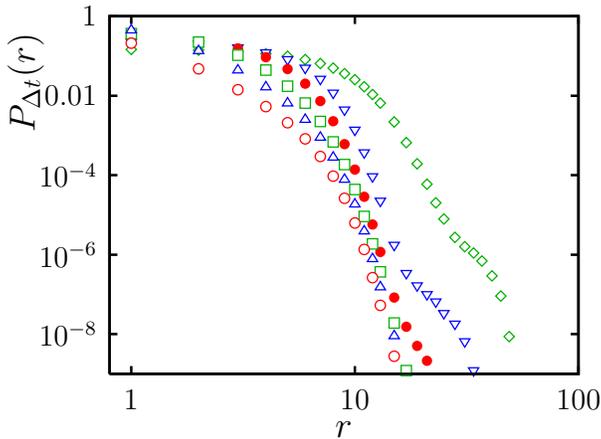}
\caption{
Distribution of returns in the Maslov model with evaporation. The
parameters are  $q=0.05$, $\overline{N}=1000$. The lime lags are
 $\Delta t=1$ ({\Large $\circ$}), $10$ ($\triangle$), $100$
($\Box$), $10^3$ ({\Large $\bullet$}), $10^4$
 ($\bigtriangledown$), and $10^5$ ({\Large $\diamond$}).
}
\label{fig:maslovmodel-m5-hist-05-allags}
\end{figure}

We can look at this behaviour from another aspect when we directly
calculate the time average\\ $\langle
v\rangle=\lim_{T\to\infty}\frac{1}{T}\sum_{t=0}^T v_t$. Its dependence
on $g$ is shown in Fig. \ref{fig:genoa-volatility}. This plot requires
some explanation. The actual implementation of the algorithm prevents
the average volatility from diverging. Instead, it reaches a
relatively large value above $10^8$. So, all points beyond this level
should be considered as effectively infinite. Moreover, in
Fig. \ref{fig:genoa-volatility} we can see a sign of bistability, or
hysteresis, which is at first sight a signature of a first-order phase
transition. However, a more careful analysis with varying $N$ shows
that the presence of an apparent hysteresis curve is
misleading. Actually, it is a subtle finite-size effect and the
phase transition is continuous (i. e. second order).

 We can see that
the transition points found independently in
Figs. \ref{fig:genoa-return} and \ref{fig:genoa-volatility} are
consistent, so it is indeed a single transition with two aspects. In
fact, the coincidence between Figs. \ref{fig:genoa-return} and
\ref{fig:genoa-volatility} means equality of time and ``ensemble''
averages, i. e. ergodicity of the model dynamics.

In Fig. \ref{fig:genoa-phasediagram} we show a phase diagram of the model,
indicating the dependence of the critical point $g_c$ on the parameter
$b$. When $b$ approaches its lower limit equal to $2$ (note that there
are no trades for $b<2$), the critical value $g_c$ diverges. It comes
as no big surprise, because trades became more rare when $b\to 2$ and
therefore the volatility diminishes. This allows the feedback measured
by $g$ to be stronger without divergence in the realised average
volatility. The phase diagram depends on the maximum number of orders
$N$, but we found that the dependence is very weak and never changes
the qualitative look of the phase diagram. The reason for this is that
for large $N$ the actual number of orders present in the system is
maintained mainly by the annihilation by other orders and the fraction
of orders which live long enough to be discarded at the end of their
lifetime is very small. In other words, the average number of orders
in the system $\langle N_\mathrm{present}\rangle$ grows extremely
slowly with $N$. 

\begin{figure}[t]
\includegraphics[scale=0.95]{%
\slaninafigdir/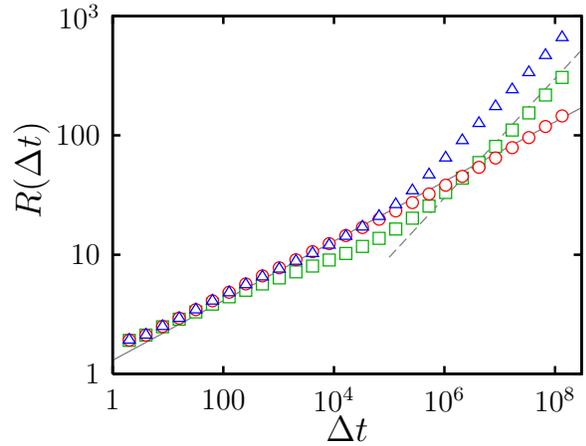}
\caption{
Hurst plot for the Maslov model without
evaporation ({\Large$\circ$}) and with evaporation probability
$q=0.01$ ($\triangle$)  and $0.05$ ($\Box$). Average number of orders
is $\overline{N}=1000$.
The solid line is the power $\propto(\Delta t)^{1/4}$,
the dashed line is $\propto(\Delta t)^{1/2}$.
}
\label{fig:maslovmodel-m5-hurst-all}
\end{figure}

To complete the study of the Genoa market model, we show in
Fig. \ref{fig:stigler-s4-autocor} the autocorrelations and in
Fig. \ref{fig:stigler-s5-s6-s7-hurst} the Hurst plot. Contrary to both
the
Stigler model and its free variant, the autocorrelation of absolute
returns decays as a clear exponential, although the characteristic
time is extremely long. As for the Hurst exponent, is is equal to
$H=1/2$, in accord with the behaviour of the free Stigler model. In
both Genoa and free Stigler models the long-time behaviour of
$R(\Delta t)$ is
dominated by the diffusion of the bunch of orders as a whole. What
makes difference between the two is the dynamics within the bunch, but
this is not visible in the Hurst plot. Note also that for the
parameters used in Fig. \ref{fig:stigler-s5-s6-s7-hurst} the regime with
$H=1/2$ starts at times $\simeq 10^5$. At such time scale the
autocorrelations are already damped out, regardless the power-law
decay in free Stigler or the slow exponential decay in Genoa models
(compare Figs. \ref{fig:stigler-s2-s3-autocor} 
and \ref{fig:stigler-s4-autocor}).

\subsection{Maslov model}
\label{sec:maslovmodel}

So far, the models investigated did not distinguish between limit
orders and market orders. The distinction was only implicit. 
All bids placed below the lowest ask acted
effectively as limit orders, as well as the asks placed above the
highest bid. In the model of Maslov \cite{maslov_99} the orders of unit
volume were issued at each step, being limit orders or market orders
with equal probability $1/2$. The limit orders were placed at close
vicinity of the current price. Here we add also the feature of order
evaporation, as in \cite{cha_sti_01}. Each order present in the book
will have the same probability of being cancelled
(evaporated). Therefore, we do not take into account the age of the
order, as we did in various variants of the Stigler model.

\begin{figure*}[t]
\includegraphics[scale=0.95]{%
\slaninafigdir/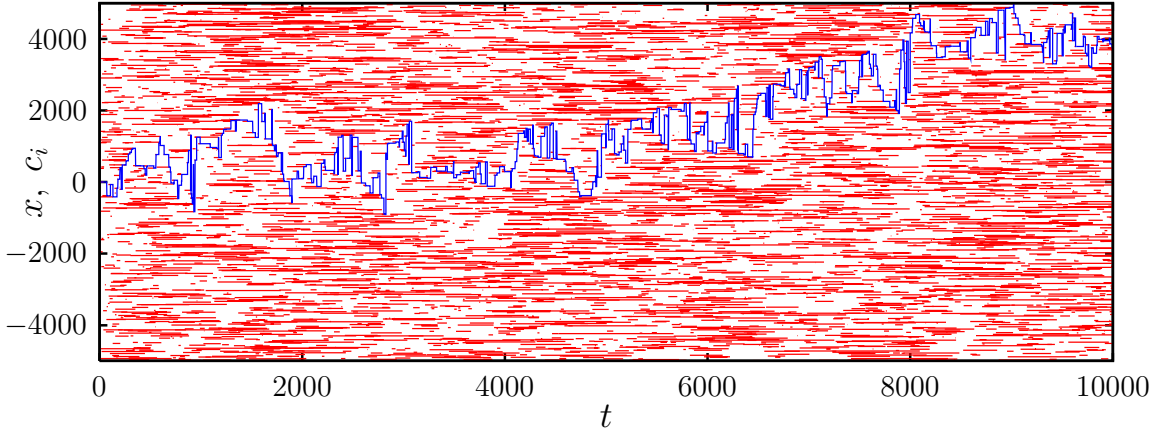}
\caption{
Example of the evolution of the Uniform Deposition Model. Each
segment of a horizontal line corresponds to one order. 
The rugged line is the time dependence of the actual price.
 The width of
the segment of allowed prices is $L=10^4$. Average
number of orders is $\overline{N}=100$ and the evaporation probability
$q=0.9$.
}
\label{fig:maslovmodel-m4-evol-1-10000-0.9-100-10000}
\end{figure*}

We tune the speed of the evaporation by a parameter $q$. For
simpler terminology, we shall call it evaporation probability. 
 Actually,
the probabilities of deposition, satisfaction and evaporation event in
one step of the evolution, at time $t$, will be defined as, respectively,
\begin{equation}
\begin{split}
&W^{+\mathrm{dep}}_t=\frac{1}{2+q\,\Big(\frac{N_t}{\overline{N}}-1\Big)}\\
&W^{-\mathrm{sat}}_t=\frac{1-q}{2+q\,\Big(\frac{N_t}{\overline{N}}-1\Big)}\\
&W^{-\mathrm{eva}}_t=\frac{q\,\frac{N_t}{\overline{N}}
}{2+q\,\Big(\frac{N_t}{\overline{N}}-1\Big)}
\end{split}
\label{eq:orderbook-maslov-depsatevaprob}
\end{equation}
where $N_t$ is the actual number of orders in the book. 
The parameter $\overline{N}$ controls the number of orders in the book
and again, to simplify the terminology, it 
will be called average number of orders,
although the actual value of the average number of orders is slightly
different (due to the effect of fluctuations). If the evaporation
probability is zero, the parameter $\overline{N}$ becomes irrelevant
for the dynamics. Note that the three probabilities
(\ref{eq:orderbook-maslov-depsatevaprob}) change in time, as the total
number of orders $N_t$ fluctuates. 

The orders are
placed at integer positions denoting the (logarithm of the) price. 
Let $x_t$ be the price at time $t$ and $N_{At}$, $N_{Bt}$ actual number
of asks and bids, respectively, with the total number of orders 
$N_t=N_{At}+N_{Bt}$. 

In case deposition is selected to
happen, according to probabilities
(\ref{eq:orderbook-maslov-depsatevaprob}), we add an ask ($N_{At+1}=
N_{At}+1$) or a bid ($N_{Bt+1}=
N_{Bt}+1$) with equal probability. The position of the new order is
$c_t=x_t+ 1$, for the ask and $c_t=x_t- 1$ sign for
the bid. The price remains unchanged, $x_{t+1}=x_t$ 
because no transaction occurred.

The execution, or satisfaction, of an order happens always when a
market order is issued, and there is a limit order to match it. Again,
sell and buy side are equivalent, so they are selected with equal
probability $1/2$. Suppose a sell order is issued and there is at
least one bid,  $N_{Bt}>0$, and $c_B$
is the position of the highest bid. Then, the new price is
$x_{t+1}=c_B$, we update $N_{Bt+1} =N_{Bt}-1$ and remove the order at $c_B$
from the book. Symmetrically it holds for the buy order.

When the evaporation of an order is about to happen, we select any of
the existing orders with uniform probability and remove it from the
system. Note that removals of a bid and an ask are not equiprobable,
as we evaporate a bid with probability $N_{Bt}/N_t$ and an ask with
probability $N_{At}/N_t$.

We can see in Fig. \ref{fig:maslovmodel-m5-evol-1-100000-0.05-100}
the space-time diagram of a typical evolution of the order book. The
price ``sows'' new orders along its fluctuating path, which are either
satisfied, as the price returns next to its original position, or they
vanish by evaporation. Longer price jumps occur when the density of
orders is low. Conversely, the price becomes temporarily pinned, when
it enters a region with large density of orders.

Let us first revisit the results for the original Maslov model without
evaporation ($q=0$).
In Fig. \ref{fig:maslovmodel-m5-hist-0-allags} we show the
distribution of returns at several time lags
\begin{equation}
P_{\Delta t}(r)=\langle\delta(r-|x_t-x_{t-\Delta t}|)\rangle\;.
\end{equation}
We can see clearly the power-law tail $P_{\Delta t}(r)\sim r^{-3}$,
observed first in \cite{maslov_99}. The results can be also rescaled
to fall onto a single curve, $P_{\Delta
  t}(r)=\frac{1}{s}F\big(\frac{r}{s}\big)$ as shown in 
Fig. \ref{fig:maslovmodel-m5-hist-0-allags-scaled}. The dependence of
the scaling factor $s$ on the time lag $\Delta t$ is shown in the
inset of Fig. \ref{fig:maslovmodel-m5-hist-0-allags-scaled} and we can
clearly see the power-law dependence $s\propto (\Delta
t)^{1/4}$. Hence we deduce the Hurst exponent of the price fluctuation
process $H=1/4$. The same value of the Hurst exponent is confirmed
independently by drawing the Hurst plot,
Fig. \ref{fig:maslovmodel-m5-hurst-all}.

The volatility clustering, measured by the autocorrelation of
absolute returns, is shown in
Fig. \ref{fig:maslovmodel-m5-autocor-all}. The autocorrelations decay
as a power law, similarly as in the Stigler model, but now the
exponent is significantly lower, $\langle|r_t\,r_{t-\Delta t}|\rangle_c\sim
(\Delta t)^{-0.5}$, which makes the behaviour much more similar to
empirical price sequences. 

Now we investigate the effect of finite evaporation probability,
$q>0$. In the distribution of one-step returns, Fig. 
\ref{fig:maslovmodel-m5-hist-all}, it $q$ leads to
deformation of the original power-law dependence. At very small values of
$q$, we observe an effective increase of the power-law exponent, to
values $1+\alpha=4$ and even more. This would sound fine, as this is
just the value reported in empirical studies. However, a cutoff starts
developing as well and when we increase $q$ further, the cutoff
prevails and the power-law regime vanishes completely. 
Since the evaporation destroys the power law, it is not surprising
that the scaling also breaks down. In
Fig. \ref{fig:maslovmodel-m5-hist-05-allags} we can see that no
scaling can be seen, because 
at each time lag the shape of the graph is different.

\begin{figure}[t]
\includegraphics[scale=0.95]{%
\slaninafigdir/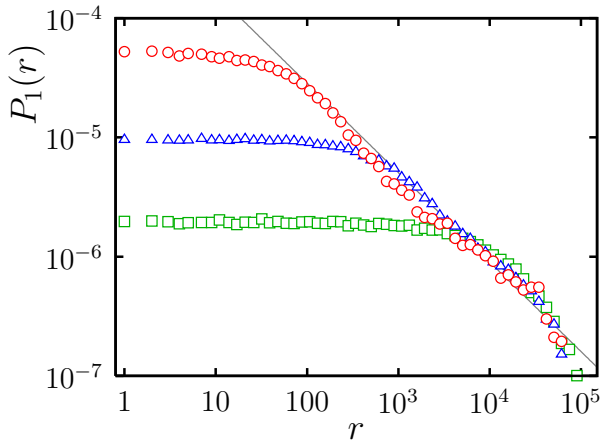}
\caption{
Distribution of one-step returns in UDM. The parameters are $L=10^6$,
$q=0.9$, and $\overline{N}=10^4$ ({\Large $\circ$}), $10^3$
($\triangle$), and $100$   ($\Box$).
The line is the power $\propto r^{-0.75}$.}
\label{fig:maslovmodel-m4-hist-9-alldensit}
\end{figure}

While the return distribution changes substantially, the absolute
return autocorrelation remains nearly the same. The decay follows
again a power law, but the exponent is somewhat larger,
 $\langle|r_t\,r_{t-\Delta t}|\rangle_c\sim
(\Delta t)^{-0.62}$. The long-time correlations are caused by the
immobile orders who sit within the book until the price finds its path
back to them. Evaporation removes some of the orders, thus eroding the
correlations. Quantitatively it results in suppression of 
the correlation function.

Finally, we look at the Hurst plot, Fig. 
\ref{fig:maslovmodel-m5-hurst-all}. As mentioned already in
\cite{cha_sti_01}, evaporation of orders induces the crossover to purely
diffusive behaviour, $H=1/2$ at large times. Interestingly, when we
compare the quantity $R(\Delta t)$ at equal time difference for
different values of $q$ we can see that larger evaporation probability
actually suppresses the diffusion. The Hurst exponent $H=1/2$ remains
universal, but the diffusion constant is lower for larger $q$. The
possible explanation is that the evaporation events go at the expense
of satisfaction events. Therefore, there are less trades per unit of
time, hence the slower diffusion of the price.

We studied also another modification of the Maslov model, where the
evaporation of orders was implemented in the sense of Stigler
model. Instead of removing an arbitrarily chosen order with fixed
probability, we track the age of the orders and remove them if the age
exceeds certain fixed lifetime. We did not observe much difference
compared to the variant with usual evaporation.  The Hurst plot looks
much like that of Fig. \ref{fig:maslovmodel-m5-hurst-all}, showing
clear crossover from the short time $H=1/4$ to long-time $H=1/2$
behaviour. Absolute returns autocorrelation decays as a power with
similar (slightly larger) exponent. Somewhat larger difference can be
seen in the return distribution. The finite lifetime of the orders
leads to decrease in the exponent of the power-law part, while the
evaporation causes its increase. Qualitatively, the 
cutoff at larger returns seems more severe than in the case of
evaporation, although quantitative comparison is hardly possible. To
sum up, we consider the variant with finite lifetime farther from the
reality than the variant with simple evaporation.

\begin{figure}[t]
\includegraphics[scale=0.95]{%
\slaninafigdir/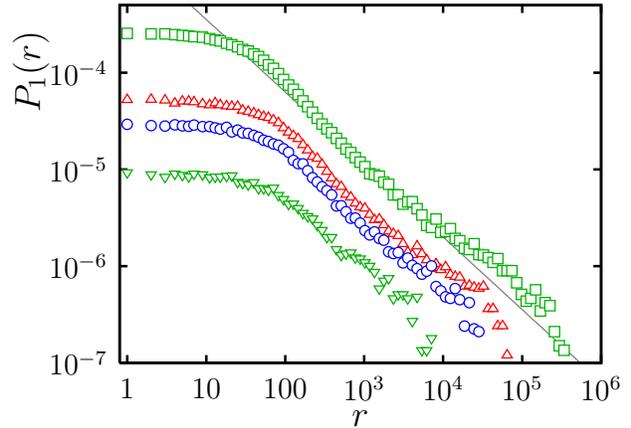}
\caption{
Distribution of one-step returns in UDM. The parameters are $L=10^6$,
 $\overline{N}=10^4$, and  $q=0.5$ ($\Box$),
 $0.9$ ($\triangle$), $0.95$ ({\Large $\circ$}), and $0.99$
($\bigtriangledown$). The line is the power $\propto r^{-0.75}$.} 
\label{fig:maslovmodel-m4-hist-allevapprob}
\end{figure}

\subsection{Uniform Deposition Model}

In Maslov model, the new orders are placed locally, at  distance $1$
from the actual price. It could be possible to fix another limit for
the maximum distance, and indeed, in the original work
\cite{maslov_99} this number was $5$. There is little, if any, effect
of the precise value of this parameter. The important thing is that
the orders are never placed farther than certain predefined limit. 

In reality, however, the distribution of distances at which the orders
are placed is rather broad and decays as a power law
\cite{bou_mez_pot_02}. The mechanism responsible for this power law is
probably related to the optimisation of investments performed by
agents working at widely dispersed time horizons
\cite{lillo_06}. Actually it is reasonable to expect that the
distribution of time horizons and (related to it) distribution of
distances is maintained by equilibration, so
that all agents expect just the same average gain, irrespectively of
the time horizon on which they act. This idea would certainly deserve
better formalisation. 

Instead of taking the empirical distribution of placements as granted
without deeper theoretical understanding, we prefer to compare the
localised deposition in Maslov model with a complementary strategy
applied in the set of models
investigated by Daniels, Farmer and others
\cite{smi_far_gil_kri_02,dan_far_ior_smi_02,dan_far_gil_ior_smi_03,ior_dan_far_gil_kri_smi_03}.
Instead of keeping 
short distance from the price, the orders are deposited with equal
probability at  arbitrary distance. In this work, we adopt one of
the variants studied in \cite{smi_far_gil_kri_02} and within this
paper we shall 
call it Uniform Deposition Model (UDM).

\begin{figure}[t]
\includegraphics[scale=0.95]{%
\slaninafigdir/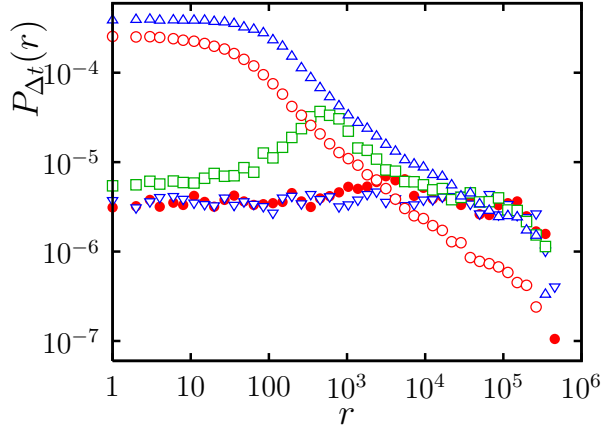}
\caption{
Distribution of  returns in UDM at different time lags.
The parameters are $L=10^6$,
 $\overline{N}=10^4$, and  $q=0.5$. 
The time lags are 
$\Delta t=1$ ({\Large $\circ$}), $10$ ($\triangle$), $100$
($\Box$), $10^3$ ({\Large $\bullet$}), and $10^4$
 ($\bigtriangledown$).  
}
\label{fig:maslovmodel-m4-hist-5-allags}
\end{figure}

In fact, the only difference with respect to the Maslov model with
evaporation, defined in Sec. \ref{sec:maslovmodel} is that we limit
the price to a segment of length $L$ and orders are deposited
uniformly on this segment. So, the orders and price can
assume integer position from the set $S=\{-L/2,-L/2+1,\ldots,L/2-2,L/2-1\}$.
As in the Maslov model, there are three classes of events, deposition,
order satisfaction, and evaporation. Their probabilities are defined
by the same formulae (\ref{eq:orderbook-maslov-depsatevaprob}) 
as in the Maslov model.  When an order is to be deposited, we first
look where is the price $x_t$. Then, select randomly a point $c_t$
from the set $S\backslash\{x_t\}$ and deposit an order there. If $c_t>x_t$ the
order becomes an ask, if $c_t<x_t$ it is a bid. (We
forbid depositing exactly at the price position.)
Although the probabilities (\ref{eq:orderbook-maslov-depsatevaprob})
look the same as in the Maslov model, we should note that there is a big
difference in the typical values of the evaporation probability
$q$. In Maslov model the orders are clustered around the price and the
evaporation is somehow a complement or correction to the natural
satisfaction of the limit orders by incoming market orders. So, $q$ is
typically a small number compared to $1$. On the contrary, in UDM the
evaporation is essential, because orders are deposited in the whole
allowed segment and ought to be removed also from areas where the
price rarely wanders. Therefore, $q$ is comparable to, although smaller
than, one. Very often, the simulations were performed in the regime
where $1-q$ was much smaller than $1$.

To see a typical situation, we plot in
Fig. \ref{fig:maslovmodel-m4-evol-1-10000-0.9-100-10000} the
space-time chart of orders and price. We can see how the price
``crawls'' through a see of orders and the configuration of the orders
changes
substantially also very far from the price and without being affected
by its movement. Of course, this is to be expected due to uniform
deposition rule. On the other hand, this is certainly not a realistic
feature. 

We found fairly interesting, although absolutely unrealistic, the
distribution of one-step returns, as shown in
Figs. \ref{fig:maslovmodel-m4-hist-9-alldensit} and
\ref{fig:maslovmodel-m4-hist-allevapprob}. The tail is characterised  by
power-law decay $P_1(r)\sim r^{-0.75}$ and the exponent, close to the
fraction $3/4$, seems to be universal, irrespectively of the
parameters $q$ and $\overline{N}$. The value of the exponent is far
below the empirical value, but the very fact of universal behaviour in
such reaction-deposition model calls for explanation. We do not have
any yet.

\begin{figure}[t]
\includegraphics[scale=0.95]{%
\slaninafigdir/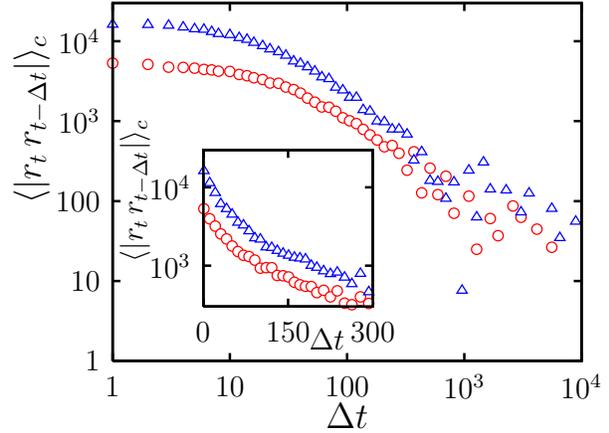}
\caption{
Autocorrelation of absolute returns in UDM. 
The parameters are $L=10^5$,  $q=0.9$; 
 $\overline{N}=10^3$ ({\Large$\circ$}), and $100$ ($\triangle$).
}
\label{fig:maslovmodel-m4-autocor-all}
\end{figure}

While the power law in the return distribution indicates some
scale-free behaviour at single time, we find no sign of scaling when
we compare the returns at different time scales. We can see that in
Fig. \ref{fig:maslovmodel-m4-hist-5-allags}. At longer lags the
power-law tail vanishes and the distribution becomes uniform. This
means that after long enough time the price can jump arbitrarily from
one position to another within nearly all the allowed range, except
 the vicinity  of the extremal points. In fact, the same behaviour was
 observed also for long enough time lags in the Stigler
 model. Certainly, the origin of such behaviour is the very existence
 of the limited price range, both in UDM and the Stigler model. 

Let us look on the volatility clustering now. In
Fig. \ref{fig:maslovmodel-m4-autocor-all} we show the autocorrelation
of absolute returns. the decay is rather slow, i. e. slower than
exponential, but at the same time it is faster than a power law. This
behaviour is special to the Uniform Deposition Model.

\begin{figure}[t]
\includegraphics[scale=0.95]{%
\slaninafigdir/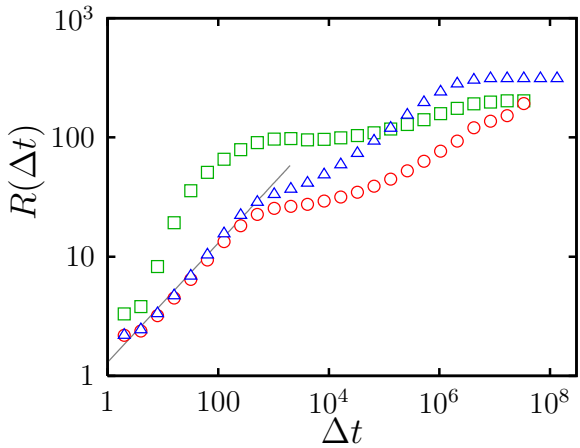}
\caption{
Hurst plot for  UDM.
The parameters are $L=10^6$;  $q=0.9$ ({\Large $\circ$}, $\triangle$), and $0.5$ 
($\Box$); 
 $\overline{N}=10^4$ ({\Large $\circ$}, $\Box$) and $1000$ ($\triangle$).
The line is the power $\propto(\Delta t)^{1/2}$.
}
\label{fig:maslovmodel-m4-hurst-all}
\end{figure}

Finally, in Fig. \ref{fig:maslovmodel-m4-hurst-all} we show the Hurst
plot. Again, there is close similarity to the Stigler model in the
sense that there is no long-time diffusive regime but saturation is
observed instead. Only in the very short initial transient we observe
ordinary diffusion-like behaviour characterised by $H=1/2$. It is
unclear from our simulations whether there is an intermediate time
window in which a non-trivial Hurst exponent (like the notorious
$H=1/4$) would be observed.

\section{Conclusions}

It is not easy to make a synoptic comparison of the whole ensemble of
models studied here. However, one easy conclusion can be drawn, that
none of them reproduces satisfactorily the reality. Most importantly,
the empirically observed Hurst exponent $H\simeq 0.6$ is not found
anywhere. 
We can classify the diffusion behaviour into three main
types. The first and most trivial one is dominated by the
saturation, $H=0$ and happens always when the price is restricted by
definition to an interval, like in the Stigler and Uniform Deposition
models. The same holds also for the asymptotic regime of the BPS
model, although in the latter the interesting things happen at the
intermediate time scale, where $H=1/4$. We do not exclude the
possibility that also in UDM the intermediate times have $H=1/4$, but
we were not able to make any conclusive statement about that.
 The second type is
characterised by asymptotic sub-diffusion, with $H=1/4$. Strictly
speaking this holds only for the Maslov model without evaporation. The
third and most frequent type of behaviour can be described as ordinary
diffusion ($H=1/2$) at long times. The initial transient regime may
exhibit either $H=1/4$, as in the Maslov model with evaporation or with
fixed finite lifetime of orders, or it may instead show the tendency to
saturation, as in the free Stigler model and Genoa artificial market model.
It seems really difficult to design an order-book model where
super-diffusive behaviour ($H>1/2$) would arise naturally, without
being put in by hand. We cannot resist the temptation to compare this
difficulty with the situation in stochastic modelling by
continuous-time random walks \cite{scalas_05}. There also, the
sub-diffusive behaviour can be found easily, but the super-diffusive
one should be essentially forced.

The power-law tails in the return distribution seem to work slightly
better. When we set apart the BPS model, where the tail decays even
faster than exponentially, we can distinguish the models where the
exponent in the power-law decay is far too low ($\alpha<0$), 
which comprises
Stigler model, free Stigler model and UDM, from the models, where the
exponent lies close, although not always precisely at the empirical
value. The latter group contains the Genoa market model and the Maslov
model with and without evaporation. The best chance for success when
matched with the real data has the Genoa model, where the exponent can
be tuned by variation of the model parameters. On the other hand, it
is a priori unclear, why the parameter values should be this and not
that. In the Maslov model proper, the exponent is universal,
$\alpha=2$. Adding evaporation increases this value, so the agreement
with the data can be again tuned, in this case by changing the
evaporation speed. However, evaporation induces not only effective
increase of the exponent, but also emergence of a cutoff. In fact, we
think that the change in exponent is only an illusion brought about by
combination of the power law and a weak cutoff. This contrasts with
the Genoa model, where, below the phase 
transition, the power-law tails are genuine for all values
of the parameter $g<g_c$. 

The very existence of the phase transition
in the Genoa market model is a remarkable fact. It is intimately
related to the dependence of the tail exponent on $g$. When the
exponent drops to the value $\alpha=1$ the average return diverges and
the transition occurs. One could speculate, how the picture would
change if the feedback between volatility and order placement was
defined differently. For example, the volatility can be defined through
squares of returns, instead of absolute returns. This would also sound 
more natural, we think. We expect that in this case the transition
would be related to the divergence of the second moment of the return
distribution, i. e. it would be located at such parameter values which would
imply the exponent $\alpha=2$. Otherwise, the picture would be most
probably the same. 

There is one feature, not so much important as such, but showing that
the free Stigler model, Genoa stock market and Maslov model are
members of the same family. If we look at the return distribution at
small returns, we find that Genoa stock market and Maslov model
(see Ref. \cite{maslov_99})
 exhibit another power-law regime, with very small
exponent $1+\alpha\simeq 0.5$. Clearly it is the sign that deep within
the bunch of orders surrounding the price 
the two models behave just like the free Stigler
model, which shows the same power law in entire range of returns. 

The return distribution in the Maslov model without evaporation has 
a very important and appealing feature. Its is the scaling
property. The returns at different time lags scale with Hurst exponent
equal to $H=1/4$. Qualitatively it agrees with the empirically found
scaling, but, unfortunately, quantitatively it is completely off. An
important finding is that the evaporation of orders destroys the
scaling, which is also absent in the UDM model. On the contrary, we
also observed scaling in the Genoa market model, but not a perfect
one. The difference between different lags is in the (not so much
important, after all) low-return range, where the power-law tail is
not yet developed.

When we want to compare the volatility clustering measured through the
autocorrelation of absolute returns, we exclude the BPS model. Due to
rather long waiting times, the measurement of the autocorrelation was
impractical. In all remaining models, we found slow decay of the
autocorrelations, but the functional form was not always a power. In
fact, there are two exceptions. In the Genoa market model, the decay
is exponential, although very slow. In UDM, the decay is faster than
any power-law but slower than an exponential. A stretched exponential
may be perhaps the candidate. In the remaining models, the power-law
decay is observed. The difference lies in the exponent. While in the
Stigler and free Stigler model, the exponent is above $1$, in the
Maslov model, both with and without evaporation, the value lies at or
close to $1/2$. 

A crucial conclusion from the above is, that we cannot simply pick a
model (``the best one'') from those studied here and apply it directly
for a stock-market practice, e. g. for option pricing. All the models
need some extensions or modifications to serve well as a realistic
description. In this work we had no intent to amend the models by
gluing together ad hoc parts with the only scope to get exponents
right. We consider that counter-productive. If a simple, bare model is
not satisfactory, one should look for another one, preferably as
simple as the first one. That is why we strove to compare ``bare''
models here. To express our feeling, the models which passed the tests
with highest scores were the Genoa market model and the Maslov model,
with some (but not too much) evaporation of orders. We must also note
that the empirical model of Ref. \cite{mik_far_07} reproduces the data
for return distribution by far the best accuracy. At the same time,
though, it makes use of several empirical inputs, rather than clear
microscopic mechanisms, and therefore follows somewhat different
modelling philosophy than ours. 
That is why we leave this model aside, without
neglecting its merits and importance.

To sum up, we compared several order-book models of stock-market
fluctuations. None of them is fully satisfactory yet. 
Calculating the return distribution, volatility
autocorrelation and the Hurst plot, we were able to identify which of
the models are promising candidates for future development. To tell the
names, they are the Genoa market model and the Maslov model.

\begin{acknowledgement}
This work was supported by the M\v SMT of the Czech Republic, grant no. 
1P04OCP10.001, and by the Research Program CTS MSM 0021620845. 

\end{acknowledgement}

\begin{thebibliography}{99}
\bibitem{an_ar_pi_88}
P. W. Anderson, K. J. Arrow, and D. Pines (eds.),
{\it The Economy as an Evolving Complex System}
 (Addison Wesley, Reading, 1988).

\bibitem{ma_sta_99}
R. N. Mantegna and H. E. Stanley,
{\it Introduction to Econophysics: Correlations and Complexity in Finance }
 (Cambridge University Press, Cambridge, 1999).

\bibitem{bou_pot_00}
J.-P. Bouchaud and M. Potters,
{\it Theory of Financial Risks }
 (Cambridge University Press, Cambridge, 2000).

\bibitem{schweitzer_02}
F. Schweitzer (editor),
{\it Modeling Complexity in Economic and Social Systems},
 (World Scientific, Singapore, 2002).

\bibitem{cha_zha_97}
D.~Challet and Y.-C. Zhang,
Physica A
 {\bf 246},
 407
 (1997).

\bibitem{ma_sta_95}
R.~N.~Mantegna and H.~E.~Stanley,
Nature
 {\bf 376},
 46
 (1995).

\bibitem{li_ci_me_pe_sta_97}
Y.~Liu, P.~Cizeau, M.~Meyer, C.-K.~Peng, and H.~E.~Stanley,
Physica A
 {\bf 245},
 437
 (1997).

\bibitem{go_me_am_sta_98}
P. Gopikrishnan, M. Meyer, L. A. N. Amaral, and H. E. Stanley,
Eur. Phys. J. B
 {\bf 3},
 139
 (1998).

\bibitem{ple_gop_am_mey_sta_99}
V. Plerou, P. Gopikrishnan, L. A. N. Amaral, M. Meyer, and H. E. Stanley,
Phys. Rev. E
 {\bf 60},
 6519
 (1999).

\bibitem{va_au_98}
N. Vandewalle and M. Ausloos,
Eur. Phys. J. B
 {\bf 4},
 257
 (1998).

\bibitem{bac_del_muz_01}
E.\ Bacry, J.\ Delour, and J.\ F.\ Muzy,
Phys. Rev. E
 {\bf 64},
 026103
 (2001).

\bibitem{eis_ker_04}
Z. Eisler and J. Kert\'esz,
Physica A
 {\bf 343},
 603
 (2004).

\bibitem{liu_dim_lux_07}
R.\ Liu, T.\ Di Matteo, T.\ Lux,
arXiv:0704.1338
 (2007).

\bibitem{bou_mat_pot_01}
J.-P.\ Bouchaud, A.\ Matacz, and M.\ Potters,
Phys. Rev. Lett.
 {\bf 87},
 228701
 (2001).

\bibitem{lil_mik_far_05}
F.\ Lillo, S.\ Mike, and J.\ D.\ Farmer,
Phys. Rev. E
 {\bf 71},
 066122
 (2005).

\bibitem{bia_hil_spa_95}
B.\ Biais, P.\ Hillion, and C.\ Spatt,
The Journal of Finance
 {\bf 50},
 1655
 (1995).

\bibitem{mas_mil_01}
S. Maslov and M. Mills,
Physica A
 {\bf 299},
 234
 (2001).

\bibitem{cha_sti_01}
D. Challet and R. Stinchcombe,
Physica A
 {\bf 300},
 285
 (2001).

\bibitem{cha_sti_03}
D. Challet  and R. Stinchcombe,
Physica A
 {\bf 324},
 141
 (2003).

\bibitem{gab_gop_ple_sta_03}
X. Gabaix, P. Gopikrishnan, V. Plerou, and H. E. Stanley,
Nature
 {\bf 423},
 267
 (2003).

\bibitem{ple_gop_gab_sta_04}
V. Plerou, P. Gopikrishnan, X. Gabaix, and H. E. Stanley,
Quant. Finance
 {\bf 4},
 C11
 (2004).

\bibitem{ple_gop_sta_05}
V. Plerou, P. Gopikrishnan, and H. E. Stanley,
Phys. Rev. E
 {\bf 71},
 046131
 (2005).

\bibitem{rosenow_02}
B. Rosenow,
Int. J. Mod. Phys. C
 {\bf 13},
 419
 (2002).

\bibitem{web_ros_03}
P. Weber and B. Rosenow,
Quant. Finance
 {\bf 5},
 357
 (2005).

\bibitem{web_ros_04}
P. Weber and B. Rosenow,
Quant. Finance
 {\bf 6},
 7
 (2006).

\bibitem{bou_mez_pot_02}
J.-P. Bouchaud, M. M\'ezard, and M. Potters,
Quantitative Finance
 {\bf 2},
 251
 (2002).

\bibitem{pot_bou_02}
M. Potters and J.-P. Bouchaud,
Physica A
 {\bf 324},
 133
 (2003).

\bibitem{bou_gef_pot_wya_03}
J.-P. Bouchaud, Y. Gefen, M. Potters, and M. Wyart,
Quant. Finance
 {\bf 4},
 176
 (2004).

\bibitem{bou_koc_pot_04}
J.-P.\ Bouchaud, J.\ Kockelkoren, and M.\ Potters,
Quant. Finance
 {\bf 6},
 115
 (2006).

\bibitem{wya_bou_koc_pot_vet_06}
M.\ Wyart, J.-P.\ Bouchaud, J.\ Kockelkoren, M.\ Potters, and M.\ Vettorazzo,
physics/0603084
 (2006).

\bibitem{zov_far_02}
I. Zovko and J. D. Farmer,
Quant. Finance
 {\bf 2},
 387
 (2002).

\bibitem{lil_far_man_02}
F.\ Lillo, J.\ D.\ Farmer, and R.\ N.\ Mantegna,
cond-mat/0207428.

\bibitem{lil_far_man_03}
F. Lillo, J. D. Farmer, and R. N. Mantegna,
Nature
 {\bf 421},
 129
 (2003).

\bibitem{far_gil_lil_mik_sen_03}
J. D. Farmer, L. Gillemot, F. Lillo, S. Mike, and A. Sen,
Quant. Finance
 {\bf 4},
 383
 (2004).

\bibitem{far_pat_zov_03}
J. D. Farmer, P. Patelli, and I. I. Zovko,
Proc. Natl. Acad. Sci. U.S.A.
 {\bf 102},
 2254
 (2005).

\bibitem{pon_lil_man_06}
A.\ Ponzi, F.\ Lillo, and R.\ N.\ Mantegna,
physics/0608032
 (2006).

\bibitem{far_zam_06}
J.\ D.\ Farmer and N.\ Zamani,
Eur. Phys. J. B
 {\bf 55},
 189
 (2007).

\bibitem{far_ger_lil_mik_06}
J.\ D.\ Farmer, A.\ Gerig, F.\ Lillo, and S.\ Mike,
Quant. Finance
 {\bf 6},
 107
 (2006).

\bibitem{lillo_06}
F.\ Lillo,
Eur. Phys. J. B
 {\bf 55},
 453
 (2007).

\bibitem{sca_kai_kir_hub_ted_06}
E.\ Scalas, T.\ Kaizoji, M.\ Kirchler, J.\ Huber, and A.\ Tedeschi,
Physica A
 {\bf 366},
 463
 (2006).

\bibitem{far_lil_04}
J. D. Farmer and F. Lillo,
Quant. Finance
 {\bf 4},
 C7
 (2004).

\bibitem{zhang_99}
Y.-C. Zhang,
Physica A
 {\bf 269},
 30
 (1999).

\bibitem{lo_mac_zha_02}
A.\ W.\ Lo, A.\ C.\ MacKinlay, and J.\ Zhang,
Journal of Financial Economics
 {\bf 65},
 31
 (2002).

\bibitem{gab_gop_ple_sta_03a}
X. Gabaix, P. Gopikrishnan, V. Plerou, and H. E. Stanley,
Physica A
 {\bf 324},
 1
 (2003).

\bibitem{stigler_64}
G. J. Stigler,
The Journal of Business
 {\bf 37},
 117
 (1964).

\bibitem{god_sun_93}
D.\ K.\ Gode and S.\ Sunder,
The Journal of Political Economy
 {\bf 101},
 119
 (1993).

\bibitem{ba_pa_shu_97}
P. Bak, M. Paczuski, and M. Shubik,
Physica A
 {\bf 246},
 430
 (1997).

\bibitem{krapivsky_95}
P.\ L.\ Krapivsky,
Phys. Rev. E
 {\bf 51},
 4774
 (1995).

\bibitem{bar_how_car_96}
G.\ T.\ Barkema, M.\ J.\ Howard, and J.\ L.\ Cardy,
Phys. Rev. E
 {\bf 53},
 2017
 (1996).

\bibitem{eli_ko_98}
D. Eliezer and I. I. Kogan,
cond-mat/9808240.

\bibitem{ta_tia_99}
L.-H. Tang and G.-S. Tian,
Physica A
 {\bf 264},
 543
 (1999).

\bibitem{rab_cin_foc_mar_01}
M. Raberto, S. Cincitti, S. M. Focardi, and M. Marchesi,
Physica A
 {\bf 299},
 319
 (2001).

\bibitem{cin_foc_mar_rab_03}
S.\ Cincotti, S.\ M.\ Focardi, M.\ Marchesi, and M.\ Raberto,
Physica A
 {\bf 324},
 227
 (2003).

\bibitem{rab_cin_05}
M.\ Raberto and S.\ Cincotti,
Physica A
 {\bf 355},
 34
 (2005).

\bibitem{rab_cin_foc_mar_05}
M.\ Raberto, S.\ Cincotti, S.\ M.\ Focardi, and M.\ Marchesi,
Computational Economics
 {\bf 22},
 255
 (2003).

\bibitem{rab_cin_dos_foc_mar_05}
M.\ Raberto, S.\ Cincotti, C.\ Dose, S.\ M.\ Focardi, and M.\ Marchesi,
in: {\it Nonlinear Dynamics and Heterogeneous Interacting Agents}, eds. T.\ Lux, S.\ Reitz, and E.\ Samanidou
 305
 (Springer, Berlin, 2005).

\bibitem{cin_foc_pon_rab_sca_06}
S.\ Cincotti, S.\ M.\ Focardi, L.\ Ponta, M.\ Raberto, and E. Scalas,
in: {\it The Complex Network of Economic Interactions}, eds. A.\ Namatame, T.\ Kaizouji, and Y.\ Aruka,
 239
 (Springer, Berlin, 2006).

\bibitem{co_bou_97}
R.~Cont and J.-P.~Bouchaud,
Macroeconomic Dynamics
 {\bf 4},
 170
 (2000).

\bibitem{maslov_99}
S. Maslov,
Physica A
 {\bf 278},
 571
 (2000).

\bibitem{slanina_01}
F. Slanina,
Phys. Rev. E
 {\bf 64},
 0561136
 (2001).

\bibitem{maslov_private}
S.\ Maslov,
private communication.

\bibitem{smi_far_gil_kri_02}
E. Smith, J. D. Farmer, L. Gillemot, and S. Krishnamurthy,
Quant. Finance
 {\bf 3},
 481
 (2003).

\bibitem{dan_far_ior_smi_02}
M. G. Daniels, J. D. Farmer, G. Iori, E. Smith,
cond-mat/0112422.

\bibitem{dan_far_gil_ior_smi_03}
M. G. Daniels, J. D. Farmer, L. Gillemot, G. Iori, and E. Smith,
Phys. Rev. Lett.
 {\bf 90},
 108102
 (2003).

\bibitem{ior_dan_far_gil_kri_smi_03}
G. Iori, M. G. Daniels, J. D. Farmer, L. Gillemot, S. Krishnamurthy, and E. Smith,
Physica A
 {\bf 324},
 146
 (2003).

\bibitem{mik_far_07}
S. Mike and J. D. Farmer,
Journal of Economic Dynamics and Control
 {\bf 32},
 200
 (2008).

\bibitem{cha_sti_02}
D. Challet and R. Stinchcombe,
cond-mat/0208025.

\bibitem{stinchcombe_06}
R. Stinchcombe,
in: {\it Econophysics and Sociophysics: Trends and Perspectives}, ed. B.\ K.\ Chakrabarti, A.\ Chakraborti, and A.\ Chatterjee,
 pp. 35-63
 (Wiley-VCH, Weinheim, 2006).

\bibitem{ha_zha_95}
T. Halpin-Healy and Y.-C. Zhang,
Phys. Rep.
 {\bf 254},
 215
 (1995).

\bibitem{wil_sch_cha_02}
R. D. Willmann, G. M. Sch\"utz, and D. Challet,
Physica A
 {\bf 316},
 430
 (2002).

\bibitem{osborne_65}
M. F. M. Osborne,
Econometrica
 {\bf 33},
 88
 (1965).

\bibitem{kul_ker_01}
L. Kullmann and J. Kert\'esz,
Physica A
 {\bf 299},
 234
 (2001).

\bibitem{fra_mar_mat_01}
F. Franci, R. Marschinski, and L. Matassini,
Physica A
 {\bf 294},
 213
 (2001).

\bibitem{mat_fra_01a}
L. Matassini and F. Franci,
Physica A
 {\bf 289},
 526
 (2001).

\bibitem{far_josh_00}
J. D. Farmer and S. Joshi,
Journal of Economic Behavior and Organization
 {\bf 49},
 149
 (2002).

\bibitem{challet_06}
D.\ Challet,
physics/0608013
 (2006).

\bibitem{mu_sla_sol_03}
L. Muchnik, F. Slanina, and S. Solomon,
Physica A
 {\bf 330},
 232
 (2003).

\bibitem{svo_sla_07}
A. Svoren\v c\'{\i}k and F. Slanina,
Eur. Phys. J. B
 {\bf 57},
 453
 (2007).

\bibitem{scalas_05}
E.\ Scalas,
in: {\it The Complex Network of Economic Interactions}, eds. A.\ Namatame, T.\ Kaizouji, and Y.\ Aruka,
 3
 (Springer, Berlin, 2006).

%
%
\end{thebibliography}
\end{document}